\documentclass[11pt]{article}
\usepackage[margin=1in]{geometry}
\usepackage{amsmath,amssymb}
\usepackage{natbib}
\usepackage{booktabs}
\usepackage{array}
\usepackage{graphicx}
\graphicspath{{figures/}}
\usepackage[hyphens]{url}
\usepackage{hyperref}
\hypersetup{hidelinks}

\newenvironment{keywords}{\par\medskip\noindent\textbf{Key words:} }{\par\medskip}

\title{Cluster on the Subject, Not the Record: Confidence Intervals and
Simultaneous Bands for Additive-Hazards Sequential Trial Emulation}

\author{M.\ Ehsan Karim\\[4pt]
\normalsize
\begin{minipage}{0.86\textwidth}\centering
School of Population and Public Health, University of British Columbia,
Vancouver, British Columbia, Canada; and Centre for Advancing Health Outcomes,
St.\ Paul's Hospital, Vancouver, British Columbia, Canada\\[3pt]
\texttt{ehsan.karim@ubc.ca}\\[3pt]
ORCID: \href{https://orcid.org/0000-0002-0346-2871}{0000-0002-0346-2871}
\end{minipage}}
\date{}

\begin{document}
\maketitle

\begin{abstract}
Sequential trial emulation (STE) estimates the effect of a sustained treatment by
stacking nested emulated trials with inverse-probability weighting. Additive-hazards
STE estimators of the marginal risk difference recommend the nonparametric bootstrap
without evaluating its coverage. Using a correctly-specifiable mechanism (up to a small, disclosed residual), we
compare analytic and bootstrap standard errors for both estimands.
Exploiting the closed-form linearity of the additive-hazards estimating equation, we
derive the influence functions and prove the default row-level robust variance
inconsistent for the marginal risk-difference curve, and show the same failure
empirically for the constant hazard difference: the row-level variance omits a
within-subject cross-trial covariance that is positive under a sign condition we
verify across our mechanisms, and is anticonservative at every horizon except the
first---only there, where the covariance is zero, is the row-level standard error
unimpaired. The subject-clustered variance and
multiplier bootstrap are consistent for the fixed-weight linearisation and support
simultaneous confidence bands, whose measured coverage is $0.88$.
Across our simulations the model-based and row-level robust intervals are
anticonservative and worsen with sample size, coverage falling to $0.71$ at $n=5000$;
clustering leaves the constant-hazard-difference coverage near $0.86$ at $n=5000$, and
the multiplier bootstrap leaves the risk-difference-curve coverage near $0.90$ ($0.86$
at the longest horizon). The STE constant hazard difference is a design-weighted summary
of a time-varying effect, dependent on the trial structure. We illustrate on the
Stanford heart transplant data and provide them in the steCI R package.
\end{abstract}

\begin{keywords}
Additive hazards; Confidence bands; Influence function; Marginal structural
model; Multiplier bootstrap; Robust variance; Sequential trial emulation;
Target trial.
\end{keywords}

\section{Introduction}\label{introduction}

\textbf{Background and motivation.}
Estimating the causal effect of a time-varying treatment in the presence of
time-varying confounding is a central problem in pharmacoepidemiology and
comparative effectiveness research. The target trial framework
\citep{hernan2016} formalises the emulation of a hypothetical randomised trial
from observational data; when eligibility recurs over follow-up, \emph{sequential}
trial emulation \citep{gran2010,danaei2013} emulates a trial at each eligible
time origin and pools them, recovering efficiency when initiators and events are
scarce within any single origin.

For a survival outcome, \citet{keogh2023} show that the sequential-trials
approach and inverse-probability-weighted marginal structural models target the
same marginal estimand, and advocate reporting a \emph{marginal risk difference}
rather than a hazard ratio, whose non-collapsibility and built-in selection
complicate causal interpretation \citep{hernan2010hazards,aalen2015,martinussen2013}.
They estimate the risk difference by standardising a weighted Aalen additive
hazards model, and---because the same individual contributes to several stacked
trials and the weights are estimated---recommend obtaining confidence intervals
by nonparametric bootstrap. The coverage of that procedure is never
assessed: their simulations report only bias and efficiency.

\textbf{The open problem.}
The variance-estimation question was taken up by \citet{limozin2025}, who
compared the simple sandwich estimator, the nonparametric bootstrap, the
linearised estimating-function (LEF) bootstrap \citep[a linearised form of the
estimating-function bootstrap of][]{hukalb2000}, and the jackknife for the marginal risk difference
in STE. They found the LEF bootstrap best in the small-sample, low-event,
low-prevalence regimes that motivate STE, and the sandwich estimator best and
cheapest at large samples---but their outcome model is a \emph{pooled logistic}
(discrete-time hazard) marginal structural model, and their discussion states
that ``inference procedures for STE based on additive hazard models warrant
further research.''

\textbf{Contributions.}
This paper provides exactly that, along four lines.

\textbf{Theory.} We derive the influence functions of the weighted-Aalen STE estimator
and prove three results. (i)~The fixed-weight, fixed-standardisation linearisation is asymptotically
normal, with a subject-clustered variance that is consistent for that same
linearisation (proved for the risk-difference curve E2 in Proposition~1; for the
scalar hazard difference E1 the underlying subject-level linearity is imported from
\citet{linying1994} and the result illustrated empirically); the weight-estimation
correction is derived and measured, and is empirically small. (ii)~The row-level robust standard error---the default in
\texttt{timereg}---is \emph{inconsistent}: it converges to the correct variance minus a
within-subject cross-trial covariance that is positive (hence anticonservative) under
the explicit sign condition~(C5), which we verify in all $81$ scenarios for the hazard
difference and in all $324$ scenario-horizon cells with $\tau\ge2$, and which fails only
at $\tau=1$, where it fails by construction, so its under-coverage does not vanish as the
sample grows. (iii)~A multiplier bootstrap and its simultaneous confidence bands are
asymptotically valid \emph{for this estimator}. The construction in (iii) is standard---it is the subject-level
multiplier and studentised supremum of \citet{martinussenscheike2006} and
\citet{yincai2004}, and we claim no novelty for it; what we establish is its validity
for an influence-function process that additionally carries estimated
artificial-censoring weights, clustering across nested overlapping trials, and an
empirical standardisation population. The
engine of these results is that the additive-hazards estimating equation
is \emph{linear} in the counting-process increments
\citep{aalen1989,linying1994,yincai2004}, which yields the influence function in closed form
and makes the multiplier bootstrap a direct resample of it rather than a
re-fitting or one-step linearisation---an advantage unavailable in the
pooled-logistic setting of \citet{limozin2025}.

\textbf{Practical recommendations.} A fast multiplier/LEF bootstrap with simultaneous bands whose
coverage approaches the nonparametric bootstrap's at a fraction of the cost, and clustering on
subject as the simple fix for the scalar hazard difference.

\textbf{Simulation.} A comprehensive
simulation ($81$ scenarios) that confirms the theory and quantifies every method
except the delete-one-subject jackknife (M5), whose $O(n)$ refits per replicate are
prohibitive at $n=5000$ (\S\ref{coverage-e1}) and which is instead reported only for
the applied example (Table~\ref{tab:heart}). Because the additive marginal structural model is
correctly specifiable under the additive-hazard data-generating mechanism---up to
a small residual bias at the longest horizons, which a population-limit analysis
(Web Appendix~E) traces to the baseline-only outcome adjustment rather than to the origin pooling
of $B(\cdot)$ (\S\ref{limitations})---our study isolates variance-estimation
performance from outcome-model misspecification to a degree the pooled-logistic
setting could not.

\textbf{Software.} An open-source R package, \texttt{steCI}, implementing the estimator, all
interval methods, and the study.

\textbf{Roadmap.}
Section~\ref{methods} sets up the estimator, the two estimands, and the six
confidence-interval methods (summarised in Table~\ref{tab:roadmap}).
Section~\ref{sec:theory} states the asymptotic theory---asymptotic normality of the
fixed-weight linearisation and the consistency of the subject-clustered variance
for it, the anticonservatism of the
row-level sandwich, and the validity of the multiplier bootstrap and its
bands---whose proofs are in Web Appendix~C. Section~\ref{results} reports the
simulation, Section~\ref{application} the Stanford heart transplant illustration,
and Section~\ref{discussion} the practical guidance (\S\ref{practical-guidance}),
limitations, and extensions.
Full derivations (Web Appendix~A), the weight-estimation correction (Web
Appendix~B), and a two-confounder robustness study (Web Appendix~D) are in the
Supplementary Material.

\section{Methods}\label{methods}
\subsection{Setting, estimator and estimands}\label{setting-estimator-and-estimands}
Consider $K$ visits $k=0,\dots,K-1$ (with $K=5$ in the simulation), a monotone
binary treatment $A_k$, a time-varying confounder $L_k$, and a survival outcome
with administrative censoring at $t=K$. From long-format data we construct emulated trials with
origins $0,\dots,K-1$; each later trial is restricted to individuals untreated
at trial entry, among whom those who initiate treatment at the origin form the
treated arm and those remaining untreated the comparator arm. Deviations from the
trial-baseline treatment are handled by
artificial censoring, corrected with stabilised
inverse-probability-of-artificial-censoring weights (IPACW), denoted $w_r$, estimated from two
pooled logistic models. The outcome model is a weighted Aalen additive hazards
model $dN_r(t) = Y_r(t)\,X_r'\,dB(t) + dM_r(t)$ fitted on the stacked,
weighted person-trial data, where $dN_r$ is the event-counting increment, $r$ indexes
start--stop records within
person-trials, $Y_r$ is the
at-risk indicator, $X_r$ is the (time-invariant) trial-baseline covariate vector,
$dM_r$ is the mean-zero martingale increment, and $B(\cdot)$ collects the cumulative
regression functions. Each record carries the weight $w_r$ of the fitted weighted Aalen model.

\textbf{Target estimands.} We consider two estimands. \emph{(E1) The constant hazard difference} $\gamma_A$,
the coefficient of a time-invariant treatment term. Its causal target is the
\emph{marginal} constant hazard difference of ``always treat'' versus ``never
treat'', whose true value ($-0.0566$ to $-0.0618$, computed in closed
form) is larger in magnitude than the data-generating
conditional coefficient $\alpha_A=-0.04$ because sustained treatment additionally
lowers the time-varying confounder $L$ and hence the hazard (Section~\ref{results}).
The additive hazard difference is collapsible over baseline covariates, so no
non-collapsibility bias arises in targeting it \citep{martinussen2013}; the gap
from $\alpha_A$ reflects the treatment-affected mediator $L$, not
non-collapsibility. \emph{(E2) The marginal risk
difference} $\mathrm{MRD}(\tau)=\Pr(T^{a=1}\le\tau)-\Pr(T^{a=0}\le\tau)$ at horizon $\tau$,
obtained by empirical standardisation of the cumulative regression functions
over the source-cohort visit-$0$ covariate distribution $\{L_0^{(k)}\}$ (``always
treat'' versus ``never treat'').

\subsection{Confidence interval methods}\label{confidence-interval-methods}
For E1 we compare (M1) the model-based standard error, (M2) the row-level robust
(sandwich) standard error, and (M3) a cluster-robust standard error clustering
on subject.\footnote{Fitted with \texttt{timereg::aalen(clusters = id)}, which at
$n=5000$ quantile-bins the subject clusters under its \texttt{max.clust} default
(clustering is exact at $n=300$ and $n=1000$). This binning does not materially
move E1 coverage, so the binned default reported in Table~\ref{tab:e1} ($0.855$)
stands; Web Appendix~D reports the supporting $2700$-replicate sensitivity check.} For both estimands we compare (M4) the subject-level nonparametric
bootstrap---re-forming trials, re-estimating weights and refitting the model in
each replicate \citep{keogh2023}, with percentile, basic and normal
intervals. For E1 we additionally consider (M5) the delete-one-subject jackknife. (M6) A multiplier/LEF
bootstrap for the closed-form additive-hazards estimating equation gives
pointwise analytic and multiplier intervals and simultaneous confidence bands for
the MRD curve (Web Appendix~A). We consider two versions: Approach-1 treats the
estimated weights as fixed, while Approach-2 additionally propagates
weight-estimation uncertainty by differentiating the cumulative stabilised
weights through the two logistic weight models (Web Appendix~B).
Table~\ref{tab:roadmap} summarises the six methods, the estimand each targets,
its asymptotic status, and the resulting recommendation.

\begin{table}[!htbp]\centering\small
\caption{Roadmap of the confidence-interval methods compared: the estimand each
targets, its construction, its asymptotic status---with the
Section~\ref{sec:theory} proposition that establishes it, where one applies---and
the resulting recommendation. E1 is the constant
hazard difference; E2 the marginal risk-difference curve. Coverage results for
the evaluated methods are in Tables~\ref{tab:e1}--\ref{tab:e2}; M5 was not
evaluated in the main simulation (Section~\ref{results}).}
\label{tab:roadmap}
\begin{tabular}{@{}>{\raggedright\arraybackslash}p{0.13\textwidth} >{\raggedright\arraybackslash}p{0.08\textwidth} >{\raggedright\arraybackslash}p{0.25\textwidth} >{\raggedright\arraybackslash}p{0.24\textwidth} >{\raggedright\arraybackslash}p{0.15\textwidth}@{}}
\toprule
Method & Estimand & Construction & Asymptotic status & Recommendation \\
\midrule
M1 model-based SE & E1 & Aalen model-based variance & Anticonservative (empirical; Table~\ref{tab:e1}) & Avoid \\
\addlinespace
M2 robust (row) SE & E1 & row-level sandwich; the \texttt{timereg::aalen} default & Inconsistent; anticonservative under (C5) (by the argument of Prop.~2) & Avoid \\
\addlinespace
M3 cluster-robust SE & E1 & sandwich clustered on the subject & Consistent (fixed-weight) given the subject-level linearisation imported from \citet{linying1994}; Prop.~1 is proved for E2 and is not invoked here; empirical E1 coverage $0.855$ at $n=5000$ (Table~\ref{tab:e1}) & Recommended (E1) \\
\addlinespace
M4 nonparametric bootstrap & E1, E2 & refit trials, weights and fit per resample & Valid under standard conditions (not established here); full-refit comparator & Reference \\
\addlinespace
M5 jackknife & E1 & delete-one-subject & Valid \citep[Ch.~2]{shaotu1995} & Optional \\
\addlinespace
M6 multiplier / LEF & E2 & perturb pre-computed influence functions; pointwise intervals and sup-$t$ band & Valid (Prop.~3); band simultaneous coverage $0.88$ (measured, \S\ref{coverage-e2}) & Recommended (E2) \\
\bottomrule
\end{tabular}
\end{table}

\subsection{Simulation study}\label{simulation-study}
\textbf{Design.} Following \citet{morris2019}, data are generated from the
additive-hazard marginal structural model (MSM) of \citet{keogh2021sim} across a
factorial of four factors ($81=3^4$ cells):
\begin{enumerate}
\item sample size $n\in\{300,1000,5000\}$;
\item event rate (baseline hazard $\alpha_0\in\{0.15,0.25,0.40\}$);
\item treatment prevalence ($\gamma_0\in\{0,-1,-2\}$);
\item confounding strength ($\gamma_L\in\{0.5,1.5,3.0\}$).
\end{enumerate}

\textbf{Event-rate constraint.} The event-rate range is constrained by
additive-hazard non-negativity: because the treated arm's confounder drifts
downward, its hazard $h=\alpha_0+\alpha_A+\alpha_L L+\cdots$ (at $A=1$) turns negative
once $L<-(\alpha_0+\alpha_A)/\alpha_L$, so very low baseline hazards give a degenerate
mechanism (at $\alpha_0=0.05$ about $44\%$ of hazard evaluations are truncated);
$\alpha_0\ge0.15$ keeps this negligible. We monitor the per-cell truncation fraction.

\textbf{Performance measures.} Per cell we run $1000$ replicates, with $B=200$
resamples for the bootstrap and multiplier methods, and evaluate empirical bias,
empirical standard deviation, mean estimated standard error, the ratio of the two,
empirical confidence-interval coverage and width, and construction-failure rates,
with Monte-Carlo errors. These Monte-Carlo standard errors are formed over the
$1000$ replicates and so already subsume the finite-$B$ resampling noise of the
bootstrap and multiplier methods; the sup-$t$ band's critical value is the single
$0.95$ quantile of the maximal studentised deviation, not an extreme tail, so its
$B=200$ jitter averages out across replicates.

\textbf{True values.} True values are computed in \emph{closed form}---from the two
counterfactual survival curves and the single-cohort additive-hazards estimating
equation---so the targets carry no Monte-Carlo error of their own; the derivation,
and the one negligible approximation it involves, are in Web Appendix~D.

\textbf{Target under a time-varying effect.} Because the true hazard difference under
this mechanism is \emph{not} constant---it falls from $-0.040$ at the first visit to
$-0.084$ at the fifth, as sustained treatment lowers the confounder---a ``constant
hazard difference'' is a \emph{summary} of that curve, and is identified only once the
weighting over time since initiation is declared. The single-cohort target above
weights the five intervals nearly equally. The sequential-trials estimator does not:
because a subject entering trial $g$ contributes follow-up only until the
administrative horizon, later times since initiation are supported by fewer trials,
and the stabilising numerator down-weights late control follow-up further. The
estimator therefore front-loads, and, applied to a declining curve, converges to a
target about $0.008$ smaller in magnitude than the single-cohort target. We report
bias against the single-cohort target throughout and quantify this
\emph{target-definition offset} in \S\ref{bias-of-the-point-estimator}; it is a
property of the design, not of the estimator, and it does not affect any
standard-error comparison, all of which are made against a common target.

\textbf{Computing.} The study runs as Slurm array jobs on the UBC Advanced Research
Computing (ARC) Sockeye cluster. The full data-generating process and the scenario
grid are in Web Appendix~D. Table~\ref{tab:expmap} maps every experiment and
supporting analysis in the paper---the headline study, the diagnostic decompositions,
and the robustness and sensitivity checks---to the question it addresses and the
display item that reports it.

\begin{table}[!htbp]\centering\small
\caption{Roadmap of the simulation study and supporting analyses: each
experiment, the question it addresses, and where its results appear. The
headline results are Tables~\ref{tab:e1}--\ref{tab:e2} and
Figures~\ref{fig:e1}--\ref{fig:band}; the robustness and sensitivity studies are
in Web Appendix~D.}
\label{tab:expmap}
\begin{tabular}{@{}>{\raggedright\arraybackslash}p{0.24\textwidth} >{\raggedright\arraybackslash}p{0.50\textwidth} >{\raggedright\arraybackslash}p{0.18\textwidth}@{}}
\toprule
Experiment / analysis & Question addressed, and what it shows & Location \\
\midrule
Headline factorial ($81$ cells $\times$ $1000$ reps) & \emph{Which interval methods are valid?} Model/row-robust SEs under-cover and worsen with $n$; subject-clustering and the multiplier/LEF methods recover most of the deficit but retain a residual shortfall---growing with $n$ for the E1 subject-clustered SE (and, for E2, only under strong confounding) and growing with the horizon for the E2 methods, the E2 aggregate otherwise being flat in $n$. & Tables~\ref{tab:e1}--\ref{tab:e2}; Figs.~\ref{fig:e1}--\ref{fig:band} \\
\addlinespace
Bias-eliminated coverage & \emph{Is the residual shortfall variance or bias?} It differs by estimand: for E1 recentring lifts cluster-robust coverage to near nominal (by up to $0.075$), so the residual is mostly the target-definition offset (\S\ref{bias-of-the-point-estimator}), not a variance effect; for E2 it buys at most $0.009$ averaged over scenarios, so the residual is predominantly a variance effect---with one corner of the design excepted (\S\ref{limitations}). & \S\ref{coverage-e1}; \S\ref{bias-of-the-point-estimator} \\
\addlinespace
Weight-estimation uncertainty (Approach-2) & \emph{Does propagating IPACW estimation matter?} Standard errors change by a signed median of $-0.35\%$ (up to $4.7\%$ in magnitude at late horizons), usually shrinking; the fixed-weight variance suffices in practice. & Web App.~B \\
\addlinespace
Standardisation-population variability & \emph{Does conditioning on the empirical baseline population understate variance?} No---the unconditional variant changes coverage by $\le0.001$. & Web Table~S5 \\
\addlinespace
SE-ratio limit / condition~(C5) & \emph{Is the omitted within-subject covariance share positive and stable?} For E1, $\Delta/V_{\mathrm{cl}}\approx0.43>0$ in all $81$ scenarios; for E2, positive at every $\tau\ge2$ and zero at $\tau=1$. & Web Table~S6 \\
\addlinespace
Two-confounder mechanism & \emph{Does the pattern survive a structurally different mechanism?} Yes: row under-covers, cluster corrects. & Web Table~S4 \\
\addlinespace
Misspecified outcome model; time-varying effect & \emph{Does the pattern survive outcome-model misspecification or a time-varying effect?} Yes, in both; $\Delta/V_{\mathrm{cl}}>0$ in every cell. & Web Table~S7 \\
\addlinespace
Bias-corrected accelerated (BCa) probe & \emph{Is the residual under-coverage interval shape (skewness)?} No evidence of it: in a two-cell paired probe at $n=300$, BCa matched rather than beat the percentile interval. & \S\ref{limitations} \\
\addlinespace
Stanford heart transplant & \emph{Do the methods behave the same on real data?} Row-level SEs are $24$--$33\%$ smaller than cluster/jackknife/bootstrap. & Table~\ref{tab:heart} \\
\bottomrule
\end{tabular}
\end{table}

\section{Asymptotic Theory}\label{sec:theory}
We now formalise the properties that the simulation illustrates. Let $n$ be the
number of independent subjects, with asymptotics as $n\to\infty$ and the number
of visits and the horizon grid $\mathcal{T}=\{1,\dots,5\}$ fixed. The crucial
structural feature is that the analysis records are \emph{not} independent: each
subject contributes to up to five overlapping emulated trials, and each trial is
stored as one start--stop record per follow-up interval, so a fully eligible
subject contributes up to $5+4+3+2+1=15$ records and the total number of records
is $R\le 15n$ (between about $3.7n$ and $7.2n$ across our design, depending on
the event rate). Records from the same trial are increments of that trial's
counting-process integral over disjoint time intervals and are orthogonal on the
trial's own filtration, so the dependence that matters---and the dependence the
clustered sandwich restores---is that \emph{across} the trials of one subject.
The estimator is asymptotically linear with the
influence functions of Web Appendix~A; write $\mathrm{IF}^{B}_c(\tau)$ for the
subject-$c$ influence function of the cumulative regression functions, and
$\mathrm{IF}^{B}_r(\tau)$ for the per-record influence function, so that
$\mathrm{IF}^{B}_c(\tau)=\sum_{r\in c}\mathrm{IF}^{B}_r(\tau)$. With $g(\tau)$ the
standardisation gradient, the marginal risk-difference influence function is
$g(\tau)^\top\mathrm{IF}^{B}_c(\tau)$ (Web Appendix~A).

\textbf{Standing assumptions.} Throughout this section we assume the conditions
under which the sequential-trials estimand is identified and the fitted model is
the right one: consistency and sequential exchangeability given the measured
history, positivity of continued adherence, and correct specification of the
weighted additive-hazards marginal structural model of
Section~\ref{setting-estimator-and-estimands}---including the pooling convention
that the cumulative coefficients $B(\cdot)$ are common across trial origins in
time since trial baseline. These make $d\hat M_r$ a martingale increment and
$\mathrm{MRD}(\tau)$ the counterfactual contrast the estimator centres on; they
are the identification assumptions revisited in Section~\ref{limitations}.

\textbf{Where our mechanism satisfies them.} The
identification assumptions, and correct specification of the weight
\emph{denominator}, hold by construction in our
data-generating mechanisms: the adherence probability there is logistic-linear in the
current confounder with no origin or visit dependence, which is exactly the fitted
denominator. The stabilising numerator, pooled across origins, is a working model
rather than the true conditional probability; (C2) does not require otherwise.
Origin-homogeneity of $B(\cdot)$ does not hold either: the
mechanism includes a calendar drift in the confounder and restricts entry to
subjects untreated at the origin, so the origin-$m$ intercept is mildly
$m$-dependent and pooled $B(\cdot)$ is a working convention rather than an exact
property. The propositions of this section are accordingly read at the pooled
probability limit to which the weighted-Aalen fit converges---its pseudo-true value
under this working convention---where the departure enters as a bounded, disclosed
approximation error (at most about $0.001$ from the pooling itself; Web Appendix~E)
rather than a gap in the theorems. Section~\ref{limitations} returns to this.

\textbf{Regularity conditions (C1)--(C4).}
\begin{enumerate}
\item[(C1)] Subjects are i.i.d.\ and the stabilised IPACW weights satisfy the joint
  moment condition $E\{w_r^2\|X_r\|^4\}<\infty$.
\item[(C2)] The \emph{denominator} weight model---for continued adherence given the
  current time-varying confounder---is correctly specified. The numerator is a
  stabilising factor and may be any working model that is a function of time since
  trial baseline and of trial-baseline covariates that already enter the outcome
  model; its correctness is \emph{not} required, since such a factor is predictable
  and measurable with respect to $X_r$ and so leaves $E\{w_rX_r\,dM_r\}=0$ untouched.
  Both coefficient vectors are estimated by $\sqrt{n}$-consistent, asymptotically
  linear (quasi-)maximum likelihood, converging to pseudo-true values under
  misspecification; because the Bernoulli log-likelihood Hessian does not involve the
  response, the weight-model influence function of Web Appendix~B is already of
  sandwich form and remains valid at the numerator's pseudo-true value. It is that
  asymptotic linearity, not correctness, through which (C2) controls the
  weight-estimation correction of Web Appendix~B.
\item[(C3)] The normalised per-time Gram matrix $n^{-1}D(t)$, with
  $D(t)=\sum_r w_r X_r X_r^\top Y_r(t)$, converges uniformly on $[0,K]$ to a
  continuous limit $d(t)$ with $\inf_{t\in[0,K]}\lambda_{\min}\{d(t)\}>0$, at rate
  \[\sup_{t\in[0,K]}\|n^{-1}D(t)-d(t)\|=O_p(n^{-1/2});\]
  and the
  limiting covariance of the standardised risk difference is non-degenerate with
  $\mathrm{SE}(\tau)>0$. The cumulative coefficient $B$ has bounded variation on
  $[0,K]$. This rate bounds only a route the proof does not take: the
  leave-one-out argument of Lemma~R (Web Appendix~C) obtains the Gram remainder
  without it, so it is the uniform limit $d(t)$ with $\lambda_{\min}>0$, not the
  $O_p(n^{-1/2})$ rate, that is load-bearing. On the trial-relative clock the risk
set is fullest at $t=0$---first records enter at their own trial baseline---so $d(t)$
is non-degenerate throughout $[0,K]$ and $\lambda_{\min}\{d(t)\}$ is bounded below;
continuity is required only of the limit $d$, the finite-$n$ Gram being a step
function whose visit-boundary jumps are handled by the summation-by-parts of Lemma~S.
\item[(C4)] The baseline covariates admit a finite moment generating function on a
  neighbourhood of the origin wide enough to cover the cumulative covariate
  coefficients of $B(\cdot)$ on $\mathcal{T}$ (a bounded-support covariate, or the
  Gaussian confounder of our mechanism, being sufficient), and the standardisation
  map $B\mapsto\mathrm{MRD}$ is Hadamard-differentiable with gradient $g$. This
  exponential-moment condition, not merely a finite set of polynomial moments, is
  what the g-computation expectations below require, since the additive-hazards
  survival factors are \emph{exponential} in the covariates and need not lie in
  $(0,1]$.
\end{enumerate}

\textbf{Sign condition (C5).}
\begin{enumerate}
\item[(C5)] The aggregate within-subject cross-trial covariance
  $\Delta=E\{\sum_{r\neq r'\in c}\psi_r\psi_{r'}\}$, in terms of the population row
  contributions $\psi_r$ defined under \emph{Population contributions} below---a
  scalar for E1, and $\Delta(\tau)$ for each $\tau\in\mathcal{T}$ for E2---is
  strictly positive.
\end{enumerate}

\textbf{Remarks on the conditions.} For~(C1), two features of the argument fix the
exponents. It is the \emph{joint} moment, not $E\{w_r^2\}<\infty$ together with a
moment of $X_r$ separately, that is required: the weights are functions of the
covariates, so the two are strongly dependent. And it is the fourth power of
$\|X_r\|$, not the second, that the arguments consume: the compensator part of the
martingale increment, $\int Y_rX_r'\,dB$, is itself linear in $X_r$, so each row's
influence contribution is bounded by a multiple of $w_r\|X_r\|(1+\|X_r\|)$ and its
second moment by $E\{w_r^2\|X_r\|^4\}$. The same envelope governs the uniform Gram
rate assumed in~(C3). Since $X_r$ contains an intercept, $\|X_r\|\ge1$, so this
condition subsumes $E\{w_r^2\|X_r\|^2\}<\infty$ and $E\{w_r^2\}<\infty$; when the
baseline covariates are bounded it reduces to the standard $E\{w_r^2\}<\infty$. That
it holds for the running mechanism, whose stabilised weights are unbounded, is
verified in Web Appendix~C.

Turning to~(C5): because $\psi_r$ is a fixed functional of the data-generating
mechanism, it is a condition on that mechanism, not on the realised sample. It holds
whenever a subject's per-trial contributions are positively correlated. The mechanism
is overlap in calendar time: for $\tau\ge2$ two trials of the same subject integrate
the \emph{same} counting-process martingale increments over the period they share.
Overlap is generic to the sequential-trials design, but the \emph{sign} is not
implied by it---the loadings involve mean-centred design vectors and could in
principle be oppositely signed, in which case the row-level sandwich would be
conservative. Shared baseline covariates alone are not sufficient: at $\tau=1$ the
trials share every baseline covariate yet $\Delta=0$. We therefore state~(C5) as an
assumption and verify it empirically (Web Appendix~D). Only Proposition~2 uses the
positivity asserted in~(C5).

\emph{One exception, and it is structural rather than incidental.} For E2 at
$\tau=1$ we measure $\Delta(1)/V_{\mathrm{cl}}(1)$---with $\Psi_c$ and
$V_{\mathrm{cl}}=E\{\Psi_c^2\}$ defined under \emph{Population contributions}
below---to be zero
(mean below $0.001$ in magnitude across the $81$ scenarios under each of the
aggregation conventions of Web Appendix~D), because at the first horizon each
emulated trial contributes only its own first interval, and a subject's first
intervals across trials occupy \emph{disjoint} calendar periods. (C5) therefore
fails at $\tau=1$ by construction, and Proposition~2 correspondingly predicts no
gap between the row-level and subject-clustered standard errors there---which is
what we observe (E2 coverage $0.949$ for both; Web Table~S3). For E1 and for E2 at
$\tau\ge2$ the share is positive throughout: $\ge0.29$ in all $81$ scenarios for
E1, and $\ge0.09$ in all $324$ scenario-horizon cells for E2.

\textbf{Population contributions.} For the functional under consideration write
$\psi_r$ for the \emph{population} row contribution---for E2,
$\psi_r(\tau)=g(\tau)^\top\phi_r(\tau)$, with $\phi_r(\tau)$ the limit of
$n\,\mathrm{IF}^{B}_r(\tau)$ obtained on replacing $n^{-1}D(s)$ by its (C3) limit and
$d\hat M_r$ by the true martingale increment---and put $\Psi_c=\sum_{r\in c}\psi_r$
and $V_{\mathrm{cl}}=E\{\Psi_c^2\}$. These are definitions, not assumptions; only
the positivity of $\Delta$ is assumed, in~(C5).

\textbf{Scope of the propositions.} Propositions~2 and~3 below are stated for a
generic scalar functional of the fit that is asymptotically linear with row-level
contributions $\psi_r$ and subject-level contributions
$\Psi_c=\sum_{r\in c}\psi_r$. For the marginal risk difference (E2),
$\psi_r=g(\tau)^\top\phi_r(\tau)$ (Web Appendix~A). Proposition~2, its corollary and
the multiplier argument of Proposition~3 use only the decomposition
$\Psi_c=\sum_{r\in c}\psi_r$ together with an ordinary law of large numbers, so they
apply to either estimand once the corresponding functional is known to be
asymptotically linear at the subject level. \emph{Proposition~1 is proved here for
E2 only}: its argument invokes the weighted-Aalen increment structure of
Web Appendix~A, condition~(C3) on $n^{-1}D(t)$, and the Hadamard-differentiable
standardisation $B\mapsto\mathrm{MRD}$ with gradient $g$ of~(C4), none of which has
an E1 analogue. For the constant hazard difference we take the corresponding
subject-level asymptotic linearity from \citet{linying1994} rather than reproving it
for the weighted, stacked person-trial design; Proposition~2 is applied to E1
conditionally on that, and the E1 row-versus-cluster results should be read as an
empirical illustration of the same mechanism rather than as a proved corollary.
That comparison is the software analogue of the decomposition: \texttt{timereg}'s
robust variance with and without subject clustering.

\textbf{Proposition 1 (Asymptotic normality; consistency of the clustered variance).}
Under (C1)--(C4), and treating the estimated weights as fixed, the vector
$\bigl(\sqrt{n}\{\hat B(\tau)-B(\tau)\}\bigr)_{\tau\in\mathcal{T}}$ converges in
distribution to a mean-zero Gaussian vector. Write $\mathrm{MRD}_n(\tau)$ for the
same empirical standardisation applied to the \emph{true} cumulative coefficients
$B(\tau)$ over the observed baseline covariates $\{L_0^{(k)}\}$. Then
$\sqrt{n}\{\widehat{\mathrm{MRD}}(\tau)-\mathrm{MRD}_n(\tau)\}$ is asymptotically
normal for each $\tau\in\mathcal{T}$, with limiting variance exactly the quantity
$\widehat V_{\mathrm{cl}}(\tau)$ estimates. The centring is at $\mathrm{MRD}_n$,
not at the population $\mathrm{MRD}$: the standardisation population is held fixed
at its observed value, and $\mathrm{MRD}_n(\tau)-\mathrm{MRD}(\tau)$ is itself
$O_p(n^{-1/2})$ and is not absorbed here. The subject-clustered influence-function
estimator $\widehat V_{\mathrm{cl}}(\tau)=\sum_{c=1}^{n}\{g(\tau)^\top\mathrm{IF}^{B}_c(\tau)\}^2$
consistently estimates the variance of the fixed-weight (Approach-1)
linearisation of $\widehat{\mathrm{MRD}}(\tau)$, conditional on the standardisation
population. The Wald interval
$\widehat{\mathrm{MRD}}(\tau)\pm z_{1-\alpha/2}\sqrt{\widehat V_{\mathrm{cl}}(\tau)}$
is therefore asymptotically valid \emph{for that fixed-weight linearisation}. For the
estimator as computed, and for the population marginal risk difference, two terms are
omitted from $\widehat V_{\mathrm{cl}}$ and neither is shown to be negligible here.
The first is the weight-estimation term $A(\tau)\mathrm{IF}^{\theta}_c$, which (C2)
bounds at $O_p(n^{-1/2})$---the same order as the leading term---but does not render
$o_p(n^{-1/2})$. The second is the standardisation-population term:
$\widehat{\mathrm{MRD}}$ averages over the observed baseline covariates, and
propagating that population's sampling variability adds a further $O(1/n)$
contribution, again of the same order as $\widehat V_{\mathrm{cl}}$ rather than
smaller. Its omission is what ``conditional on the standardisation population''
records. We assume the negligibility of neither. Both are instead measured, and both
are small: the weight-estimation correction shifts the standard errors by a signed median of
$-0.35\%$ (Web Appendix~B), and moving the standardisation from fixed to random leaves
coverage unchanged to three decimals across the representative cells (Web
Table~S5). Approach-2 removes the need for the first by carrying the correction
explicitly, and the random-standardisation option of \texttt{steCI} for the second.
The theorem is therefore a statement about the fixed-weight,
fixed-standardisation linearisation; the bridge from it to the population marginal
risk difference is the standardisation term, whose omission from
$\widehat V_{\mathrm{cl}}$ is anticonservative by
$\mathrm{Var}\{S_0^{(i)}(\tau)-S_1^{(i)}(\tau)\}/n$ (Web Appendix~C, remark), and
the weight-estimation term, whose sign we do not determine. Because the horizon grid $\mathcal{T}$ is finite, only
finite-dimensional convergence is required; asymptotic normality of the
standardised risk difference, centred at $\mathrm{MRD}_n$, then follows from a
multivariate central limit theorem for the i.i.d.\ population subject-level
influence functions and a first-order expansion in $\hat B(\tau)$ at the observed
standardisation population, proved in Web Appendix~C
\citep{aalen1989,vandervaart1998}. We claim no weak convergence of $\hat B$ as a
process on $[0,K]$, and none is needed: every interval and band in this paper is
formed at the five grid horizons.

\textbf{Proposition 2 (Inconsistency and anticonservatism of the row-level sandwich).}
Let $\widehat V_{\mathrm{row}}(\tau)=\sum_{r=1}^{R}\{g(\tau)^\top\mathrm{IF}^{B}_r(\tau)\}^2$
be the variance estimator that treats the $R$ analysis records as independent.
Then
\[
\widehat V_{\mathrm{row}}(\tau)=\widehat V_{\mathrm{cl}}(\tau)-\sum_{c=1}^{n}\ \sum_{r\neq r'\in c}\{g(\tau)^\top\mathrm{IF}^{B}_r(\tau)\}\{g(\tau)^\top\mathrm{IF}^{B}_{r'}(\tau)\},
\]
so it omits precisely the within-subject cross-record covariances, which by the
orthogonality of same-trial increments noted above are dominated by the
cross-trial terms. Both variance
estimators are $O_p(n^{-1})$; scaled by $n$, the omitted sum converges in
probability to $\Delta(\tau)$ and $n\widehat V_{\mathrm{cl}}(\tau)$ to
$V_{\mathrm{cl}}(\tau)$, so
\[
  \widehat V_{\mathrm{row}}(\tau)\big/\widehat V_{\mathrm{cl}}(\tau)
  \;\xrightarrow{p}\;1-\Delta(\tau)/V_{\mathrm{cl}}(\tau).
\]
The omitted piece is thus a \emph{fixed fraction} of the asymptotic variance
rather than a vanishing remainder: whenever $\Delta(\tau)\neq0$ the ratio stays
bounded away from $1$, so $\widehat V_{\mathrm{row}}$ is \emph{inconsistent} for
the asymptotic variance and the deficit does not shrink with $n$. \emph{Under~(C5)}
(positive within-subject dependence) $\Delta(\tau)>0$, so the ratio is strictly
below $1$: the row-level standard error is asymptotically too small, and its Wald
intervals under-cover. Overlap is generic to the sequential-trials design, but
condition~(C5) is a sign assumption we verify rather than derive: it is satisfied
in all $81$ scenarios for E1 and all $324$ scenario-horizon cells with $\tau\ge2$,
and fails by construction at $\tau=1$. Without it the inconsistency remains but
its direction is not guaranteed. This is the clustered
estimating-equation argument
\citep{liang1986,yincai2004}: the robust variance is consistent only when summed over
the independent units (subjects), not the correlated rows. Quantitatively, the
decomposition $\widehat V_{\mathrm{row}}=\widehat V_{\mathrm{cl}}-\Delta$ gives the
row-to-cluster standard-error ratio a closed-form limit,
\begin{equation}
  \rho(\tau)=\frac{\mathrm{SE}_{\mathrm{row}}(\tau)}{\mathrm{SE}_{\mathrm{cl}}(\tau)}
  \longrightarrow\sqrt{1-\Delta(\tau)/V_{\mathrm{cl}}(\tau)}\;<\;1,
  \label{eq:seratio}
\end{equation}
so the shortfall is governed by the cross-covariance share $\Delta/V_{\mathrm{cl}}$.
Estimated from the influence functions across the $81$ scenarios, this share is
$\Delta/V_{\mathrm{cl}}\approx0.43$ for E1 (the row-level sandwich omits about $43\%$ of
$V_{\mathrm{cl}}$, the fixed-weight subject-clustered asymptotic variance) and is \emph{positive in every one of the $81$ scenarios} (minimum
$0.29$), so condition~(C5) holds throughout for E1; for E2 it is positive at every
$\tau\ge2$ (minimum $0.09$ over the $324$ scenario-horizon cells) and zero at
$\tau=1$, for the structural reason given with~(C5). The implied $\rho\approx0.75$ is
arithmetically the row-to-cluster SE ratio behind the headline row-level SE-ratio
near $0.6$ that fails to improve with $n$ (Table~\ref{tab:e1}; Web Appendix~D).

\textbf{Proposition 3 (Bootstrap and band validity; weight negligibility).}
Under (C1)--(C4), the conditional law of the multiplier process
$\sum_{c=1}^{n}\xi_c\, g(\cdot)^\top\mathrm{IF}^{B}_c(\cdot)$ given the data
converges in probability, uniformly on $\mathcal{T}$, to the sampling law of the
fixed-weight linearisation of $\widehat{\mathrm{MRD}}(\cdot)-\mathrm{MRD}_n(\cdot)$;
the same conclusion holds for the estimator as computed when the multipliers act
on the Approach-2 influence functions of Web Appendix~B, or under the empirically
verified negligibility of the weight-estimation term. Hence the pointwise
multiplier intervals are asymptotically valid and the sup-$t$ band attains nominal
simultaneous coverage in the limit; these rest on the conditional multiplier central limit
theorem for the asymptotically linear representation
\citep{vandervaartwellner1996,kosorok2008} and the continuous-mapping theorem for
the supremum functional \citep{montielolea2019}, and
treat the estimated weights as fixed. The band construction itself is that of
\citet[][\S5.2, \S5.5]{martinussenscheike2006} and \citet{yincai2004} for additive
hazards; Proposition~3 establishes it for the sequential-trial influence-function
process, which those references do not cover. Weight estimation adds a second, two-step
term to the influence function: the complete correction is Approach-2 (Web
Appendix~B), obtained by differentiating the outcome estimating equation through
the weight-model score. With \emph{stabilised} weights this term is close to
orthogonal to the outcome estimating function \citep{shu2021,enders2018}---and we find it
changes the standard errors by a signed median of $-0.35\%$. The change is negative
in most design cells, so that propagating the estimation of stabilised weights
typically shrinks the standard error rather than inflating it, as the
near-orthogonality results above predict; it vanishes at $\tau=1$, where the
cumulative weights are one by construction, and grows with the horizon as the weights
compound. The full per-horizon and per-sample-size breakdown---including the median
absolute change, the range, the fraction of cells exceeding $1\%$ in magnitude, and
the behaviour at the longest horizons---is in Web Appendix~B.
It is therefore small relative to the
standard-error shortfall we document, and the fixed-weight Approach-1 suffices for
practical purposes, but it is not uniformly below $1\%$. We report this as an
empirical finding backed by the stabilised-weight construction rather than as an
exact first-order orthogonality theorem.

\textbf{Remark (why the closed form matters).} The weighted-Aalen estimating equation
is linear in the counting-process increments---weighted least squares with no
Newton step---so the influence function is available in \emph{closed form}, without
numerical differentiation. Because the per-time Gram matrix $D(t)$ is
data-dependent the estimator is \emph{asymptotically} linear (the remainder is
$o_p(n^{-1/2})$) rather than exactly linear; but the closed form is what lets the
multiplier bootstrap perturb the pre-computed influence function directly, a
zero-step operation, rather than through the per-draw one-step linearisation that
the pooled-logistic LEF bootstrap of \citet{limozin2025} uses. Full proofs are in Web
Appendix~C.

\section{Results}\label{results}
\textbf{Overview.} The headline is a coverage failure and a partial fix. The
model-based and row-level robust standard errors are markedly anticonservative for
both estimands; for the constant hazard difference they additionally \emph{worsen}
with sample size, coverage of nominal $95\%$ intervals falling to $0.71$--$0.72$ at
$n=5000$, whereas for the risk-difference curve the row-level shortfall is roughly
flat in $n$ (coverage near $0.86$).
Clustering on the subject, and the multiplier bootstrap for the risk-difference
curve, recover most of that deficit but leave a residual shortfall---the
constant-hazard-difference coverage near $0.86$ at $n=5000$ and the
risk-difference-curve coverage near $0.90$ (falling to $0.86$ only at the longest
horizon); the simultaneous band's whole-curve coverage is $0.88$. The rest of this
section documents these results, beginning with construction and validity
bookkeeping.

The results below are from the full design---all $81$ scenarios,
$n\in\{300,1000,5000\}$---targeting $1000$ replicates per cell, run with per-method
error handling so that a failure in one interval procedure does not discard the
replicate for the others. Construction failures are rare, and they are \emph{not}
shared equally across methods. Eight of the $81$ cells contain at least one
failure; all lie at $n=300$ or $n=1000$ and predominantly at high treatment
prevalence, where the untreated arm is depleted. Across the $405$
(scenario, horizon) cells the influence-function methods (M6) fail in $0.35\%$ of
attempted replicates and their Approach-2 variants in $0.38\%$, against $0.015\%$
for the nonparametric bootstrap and for the E1 analytic standard errors---more
than an order of magnitude apart, because the M6 construction additionally
requires a non-singular weighted Gram matrix at every event time. In the worst
single cell ($n=300$, high event rate, weak confounding, high prevalence) M6
retains $812$ of $1000$ replicates and the bootstrap $988$. Performance measures
use the constructible replicates.

\textbf{Validity checks on scale statistics.} Two validity checks are applied when forming \emph{scale} statistics
for the marginal risk difference. The first is definitional: the risk difference
is a difference of probabilities and so lies in $[-1,1]$, so an estimate outside it
is an estimator failure. The second is a finiteness heuristic---a standard error
above $0.5$ implies a nominal interval wider than the entire parameter space, the
signature of estimator breakdown in a low-overlap cell rather than a usable
dispersion. Replicates violating either bound are excluded from the empirical
standard deviation, the bias and
mean squared error, and the mean estimated standard error. They are rare. Coverage is
deliberately \emph{not} subject to these checks, since discarding replicates far
from the centre would remove exactly those that fail to cover. The constant hazard
difference has no such definitional bound and no replicate was excluded there, so
every standard-error ratio reported for E1 is computed on the full set of
replicates. Web Appendix~D reports the exact counts of excluded point estimates and
interval standard errors, their extreme values, and the unguarded scale statistics
alongside the guarded ones.

\textbf{A note on the E1 target.} The causal estimand is the \emph{marginal} constant
hazard difference of ``always treat'' versus ``never treat'', which is not the
data-generating conditional coefficient $\alpha_A=-0.04$: sustained treatment
also lowers the time-varying confounder $L$, which further lowers the hazard, so
the marginal effect is larger in magnitude. Computed in closed form
(Web Appendix~D), it is $-0.0618$, $-0.0597$ and $-0.0566$ at the low, medium and
high event rates. All E1 bias and coverage below are computed against this
marginal target.

\subsection{The E1 offset is a target definition, not estimator bias}\label{bias-of-the-point-estimator}
Against the single-cohort target, the sequential-trials estimator of the constant
hazard difference sits about $+0.007$ to $+0.009$ away from it, averaging over the
$27$ scenarios at each sample size (empirical SD
$0.063$, $0.042$ and $0.025$ at $n=300,1000,5000$). Per scenario the displacement
ranges from $+0.0007$ to $+0.0143$, so the averages conceal a roughly twentyfold
spread; it is positive in all $81$ cells. This displacement is
\emph{not} finite-sample bias, and the paper's own criterion---an inconsistency
shows itself by failing to shrink with $n$---identifies it as such: regressing
$\log|\text{bias}|$ on $\log n$ within each of the $27$ scenarios gives a mean
slope of $+0.01$ (95\% CI $-0.04$ to $+0.06$), indistinguishable from zero and
excluding both the $-1$ of an $O(n^{-1})$ bias and the $-0.5$ of an $O(n^{-1/2})$
term---the latter by $20$ standard errors---and the displacement is undiminished at
$n=5\times10^{4}$.

It is instead a consequence of summarising a non-constant curve. Sustained
treatment lowers the confounder, so the true marginal hazard difference falls from
$-0.040$ at the first visit to $-0.084$ at the fifth---a factor of $2.1$---and a
\emph{constant} hazard difference is therefore identified only once the weighting
over time since initiation is fixed. The single-cohort target of
\S\ref{setting-estimator-and-estimands} weights the five intervals nearly equally
($0.25,0.22,0.20,0.17,0.15$, which reproduce the exact target to $2\times10^{-6}$);
the estimator front-loads them, because a subject entering trial $g$ is followed
only to the administrative horizon---so late times since initiation are supported
by fewer trials---and because the stabilising numerator further down-weights late
control follow-up. Front-loading a declining curve returns a smaller magnitude,
necessarily and in the direction observed.

Three checks establish that this, and not the estimator, is the source. Imposing
the stacking weights alone moves the target by $+0.0026$, and adding the
stabilising numerator moves it to $+0.0066$---the two are cumulative, not
additive---against $+0.0078$ observed. Setting the single
data-generating coefficient that makes the hazard difference time-varying to zero,
so that every weighting yields the same target, removes the displacement entirely
($-0.0003$, $0.3$ standard errors) with the estimator left untouched, and
reversing the sign of that coefficient reverses the displacement. And re-cutting
the \emph{target's} own horizon from $\tau=5$ to $\tau=3$---with no estimator
involved at all---reproduces the entire displacement.

The practical consequence is that a constant hazard difference reported from a
sequential-trials analysis is a design-weighted summary: two analyses of the same
data with different numbers of trials, or different administrative horizons, target
different scalars. We therefore treat E2---the fully specified marginal
risk-difference curve, which carries no such ambiguity---as the primary estimand for
coverage, and report E1 as the collapsible scalar that off-the-shelf software
returns. None of the standard-error comparisons are affected: every method is scored
against a common target, so the displacement is shared and cancels from every
between-method contrast.

\textbf{Bias of the E2 point estimator.} For the risk-difference curve the point
estimator carries a bias that we report
here because it varies systematically across the design. At $\tau=5$ the mean bias
is $0.019$, $0.011$ and $0.007$ at $n=300$, $1000$ and $5000$ under moderate
confounding, but $0.029$, $0.020$ and $0.016$ under strong confounding against
$0.014$, $0.004$ and $0.002$ under weak. The gradient is a property of the
estimator, not of the target: the true risk difference takes only three values
across the design, one per event rate. Two features bound how much should be read
into it. The bias decays with sample size (pooled, as $n^{-0.33}$), and it decays at
close to the rate of the estimator's own sampling variability ($n^{-0.30}$), so the
ratio of bias to empirical SD is flat to declining rather than growing
(averaging the cellwise bias-to-SD ratio over the nine cells at each sample size:
$0.19$, $0.18$, $0.19$ under strong confounding; $0.10$, $0.03$, $0.05$ under
weak; forming instead the ratio of the averaged bias to the averaged SD gives
$0.19$, $0.18$, $0.20$ and $0.13$, $0.06$, $0.06$, with the same pattern).
Over the range studied the bias is therefore not growing relative to sampling
error. Three sample sizes cannot themselves separate a bias that vanishes from one
decaying to a small plateau; a population-limit calculation (\S\ref{limitations})
settles the question, and it is a plateau---a small, localised outcome-model
misspecification at the longest horizons, with the origin pooling contributing at
most about $0.001$, rather than a bias that fully vanishes. It is small enough that identification,
not this extrapolation, remains the basis for treating the estimator as
effectively unbiased across the rest of the design. The slow decay of the empirical SD under strong confounding is the
finite-sample signature of heavy-tailed stabilised weights---condition~(C1) holding
in the limit while the sample is still some way from it---rather than evidence
against identification.

\subsection{Coverage of the constant hazard difference (E1)}\label{coverage-e1}
\begin{figure}[!tbp]\centering
\includegraphics[width=.95\linewidth,alt={Two line panels of empirical coverage and SE ratio against sample size 300, 1000, 5000 for four methods; the model-based and row-level robust curves fall from about 0.82 to about 0.71 as n grows and their SE ratios fall from about 0.7 to about 0.6, while the cluster-robust and bootstrap curves fall from about 0.92 to about 0.86 with SE ratios near 0.8 to 0.9.}]{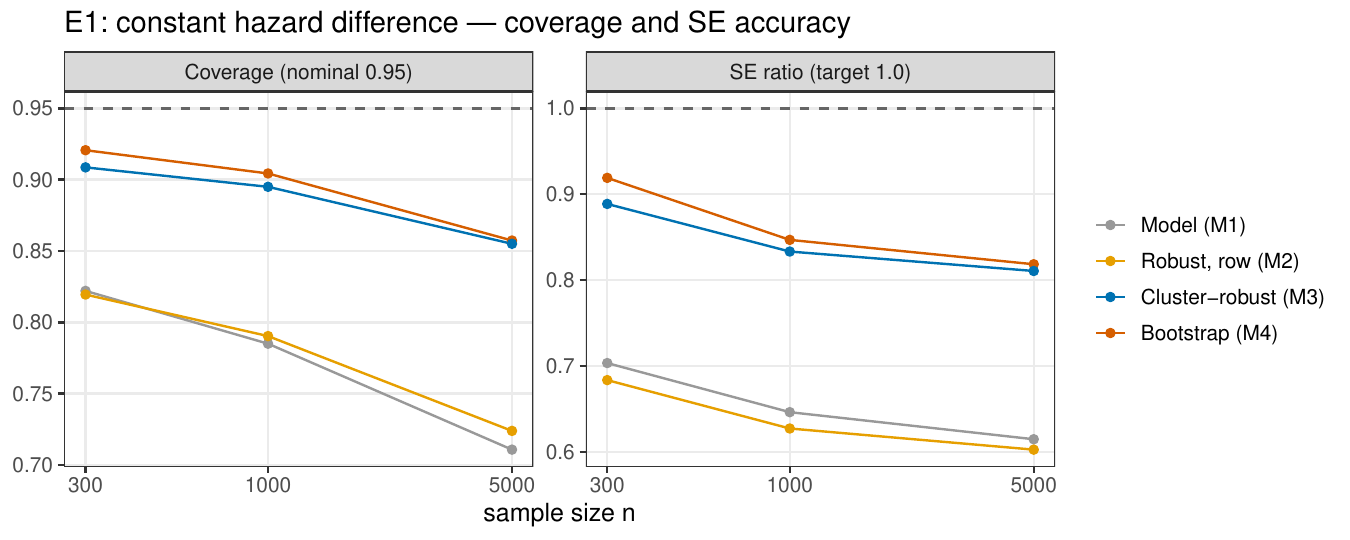}
\caption{E1 (constant hazard difference): empirical coverage of nominal $95\%$
intervals (left) and the SE ratio (right) by method and sample size, averaged
over scenarios. The model-based and row-level robust SEs under-cover
increasingly with $n$; the cluster-robust SE (M3) and the nonparametric bootstrap
(M4) recover most of the deficit. (Target $1000$ replicates per cell; $80{,}988$
of $81{,}000$ realised.)}
\label{fig:e1}
\end{figure}

Table~\ref{tab:e1} reports empirical coverage of nominal $95\%$ intervals and the
SE ratio (mean estimated SE divided by the empirical standard deviation of the
point estimates; values below $1$ indicate under-estimation of variability);
Figure~\ref{fig:e1} displays the same across sample size. Because the constant
hazard difference estimated by stacked STE is a design-weighted summary that differs
from the single-cohort target used as the truth (\S\ref{bias-of-the-point-estimator}),
part of the raw E1 shortfall below is this target-definition offset rather than a
variance-estimator failure; the bias-eliminated column Cov.$^\dagger$ isolates the
variance contribution, returning to near nominal for the cluster-robust SE (M3) while
staying low for the model-based and row-level SEs (M1, M2).

\begin{table}[!htbp]\centering
\caption{Empirical coverage of nominal $95\%$ intervals for the constant hazard
difference (E1), and the SE ratio (mean estimated SE / empirical SD), averaged
over the $27$ scenarios at each sample size (Monte-Carlo errors in \S\ref{results}). Cov.$^\dagger$ is the
\emph{bias-eliminated} coverage---of intervals recentred on the mean point
estimate, the usual recentred-coverage diagnostic---which separates the variance contribution to the shortfall from the E1
target-definition offset (\S\ref{bias-of-the-point-estimator}): where it returns to
near nominal (M3, M4) the raw shortfall was largely that offset, whereas a
Cov.$^\dagger$ that stays low (M1, M2) marks a genuine variance failure.}
\label{tab:e1}
\begin{tabular}{lccccccccc}
\toprule
 & \multicolumn{3}{c}{$n=300$} & \multicolumn{3}{c}{$n=1000$} & \multicolumn{3}{c}{$n=5000$}\\
\cmidrule(lr){2-4}\cmidrule(lr){5-7}\cmidrule(lr){8-10}
Method & Cov. & Cov.$^\dagger$ & SEr & Cov. & Cov.$^\dagger$ & SEr & Cov. & Cov.$^\dagger$ & SEr\\
\midrule
M1 model-based SE            & 0.822 & 0.845 & 0.70 & 0.785 & 0.819 & 0.65 & 0.711 & 0.808 & 0.61\\
M2 robust SE (row-level)     & 0.819 & 0.840 & 0.68 & 0.790 & 0.819 & 0.63 & 0.724 & 0.819 & 0.60\\
M3 cluster-robust SE         & 0.909 & 0.928 & 0.89 & 0.895 & 0.923 & 0.83 & 0.855 & 0.930 & 0.81\\
M4 bootstrap (percentile)    & 0.921 & 0.932 & 0.92 & 0.904 & 0.925 & 0.85 & 0.857 & 0.928 & 0.82\\
\bottomrule
\end{tabular}
\end{table}

The model-based and row-level robust standard errors---the latter being the
unclustered robust sandwich SE that \texttt{timereg::aalen} reports by default
when robust variances are requested---substantially under-cover, and, tellingly,
the under-coverage \emph{worsens} with sample size: from about $0.82$ at $n=300$
to $0.79$ at $n=1000$ and $0.71$--$0.72$ at $n=5000$, with SE ratios ranging from
about $0.70$ down to $0.60$ as $n$ grows. A proportional shortfall in the SE that does not vanish as $n$ grows is
the signature of an inconsistent variance estimator, not a finite-sample
artefact. This is consistent with the reuse of each individual across up to five
stacked emulated trials: the row-level estimators treat the resulting correlated
analysis records as independent. Clustering the robust variance on the subject
(M3)---which addresses exactly that dependence---recovers most of the deficit,
to $0.91$, $0.90$ and $0.86$ at the three sample sizes, with SE ratios near
$0.81$--$0.89$ and interval widths essentially equal to the nonparametric
bootstrap (M4). The pooled M3 SE ratio masks a systematic dependence on
confounding strength: under weak confounding it is calibrated and improves with
$n$ (about $0.99$, $0.96$, $1.00$), whereas under strong confounding---the regime
that motivates STE---it deteriorates monotonically to about $0.81$, $0.71$, $0.63$.
This is a truth-free signal, since the SE ratio never references the target; its
coverage consequence is cushioned by the heavy-tailed estimate distribution
(bias-eliminated coverage stays near $0.90$--$0.93$), so the recommendation of
M3 stands. This M3 decline differs in status from the M1/M2 shortfall:
Proposition~2 fixes the row-level limit strictly below one at every confounding
strength, whereas consistency of $V_{\mathrm{cl}}$ requires the clustered ratio to
return toward one as $n\to\infty$---which we establish for E2 (Proposition~1) but
for E1 import from \citet{linying1994} rather than prove; within our sample-size
range the strong-confounding E1 ratio does not yet turn back, an observation we
cannot presently separate from slow finite-sample convergence. A residual under-coverage nonetheless remains for M3 and widens
with $n$, from about four percentage points at $n=300$ to about ten at $n=5000$;
as the bias-eliminated coverage below shows, this residual is predominantly the
target-definition offset of \S\ref{bias-of-the-point-estimator} rather than a
failure of the clustered variance. The weight-estimation contribution to it is
examined below and with the Approach-2 variant of M6. The
bootstrap attains similar coverage, its normal and percentile variants
outperforming the basic/pivot interval. The delete-one-subject jackknife (M5) is
asymptotically equivalent to the influence-function/cluster-robust variance for a
smooth linear functional \citep{shaotu1995}, so it is expected to track M3/M4. It
was not evaluated in the main simulation---its $O(n)$ refits per replicate are
prohibitive at $n=5000$---and is therefore not reported in Table~\ref{tab:e1}.

The residual is only partly a variance effect. Bias-eliminated coverage
(intervals recentred on the mean estimate) is close to nominal for the
cluster-robust SE at all sizes---$0.928$, $0.923$, $0.930$---so at $n=5000$ the
raw M3 shortfall ($0.95-0.855=0.095$) is predominantly small-bias---recentring
lifts coverage by $0.075$---with a smaller residual SE-underestimation of about
$0.020$. For the model and row-level robust SEs, by contrast,
bias-elimination barely helps (e.g.\ row-level robust at $n=1000$: $0.790$ raw
versus $0.819$ recentred), confirming their failure is one of variance
estimation.

Under-coverage of the model/robust SEs is worst in the regimes that most stress
the method: averaged over sample sizes it falls to about $0.71$ at the low
event rate (against $0.83$ at the high rate). It is, however, \emph{milder} at low
treatment prevalence ($0.80$ for the model-based SE, against $0.75$ at high
prevalence)---a direction we report as measured, since the natural explanations,
sparser data and a more depleted treated arm at low prevalence, would predict the
opposite.
Even the cluster-robust SE drops to about $0.84$ at the low event rate while
staying near nominal ($0.90$--$0.92$) at medium-to-high rates. The breakdown by
event rate and sample size is in Table~\ref{tab:e1_eventrate}; the same breakdown,
shown graphically and split by the four E1 interval methods, is Web Figure~S1 (Web
Appendix~D).

\begin{table}[!htbp]\centering
\caption{Coverage of nominal $95\%$ intervals for the constant hazard difference
(E1) by event rate and sample size, promoting the Web Appendix~D breakdown (Web
Figure~S1). Each cell averages the nine scenarios sharing that event rate and $n$.
The anticonservatism of the model-based and row-level robust standard errors is
worst at the low event rate and \emph{deepens} with $n$ (model-based coverage
$0.78\to0.60$); the subject-clustered and bootstrap intervals recover most of the
deficit but, at the low event rate and $n=5000$, still reach only about $0.78$.}
\label{tab:e1_eventrate}
\begin{tabular}{llcccc}
\toprule
Event rate & $n$ & M1 model & M2 robust & M3 cluster & M4 bootstrap\\
\midrule
Low    & $300$  & 0.781 & 0.781 & 0.887 & 0.904\\
       & $1000$ & 0.734 & 0.742 & 0.865 & 0.877\\
       & $5000$ & 0.604 & 0.617 & 0.775 & 0.784\\
\addlinespace
Medium & $300$  & 0.823 & 0.821 & 0.910 & 0.922\\
       & $1000$ & 0.793 & 0.797 & 0.904 & 0.914\\
       & $5000$ & 0.723 & 0.736 & 0.872 & 0.874\\
\addlinespace
High   & $300$  & 0.862 & 0.856 & 0.929 & 0.936\\
       & $1000$ & 0.829 & 0.831 & 0.916 & 0.923\\
       & $5000$ & 0.806 & 0.818 & 0.918 & 0.914\\
\bottomrule
\end{tabular}
\end{table}

\subsection{Coverage of the marginal risk difference (E2)}\label{coverage-e2}
\begin{figure}[!tbp]\centering
\includegraphics[width=.9\linewidth,alt={Three panels, one per sample size (n=300, 1000, 5000), of empirical coverage of the E2 risk-difference methods against horizon one to five. Coverage declines with horizon for the pointwise methods: the cluster-robust analytic and multiplier curves fall from about 0.94 toward 0.88 and the row-level analytic curve lower still, while the simultaneous band sits above the rest; the declining pattern holds at every sample size.}]{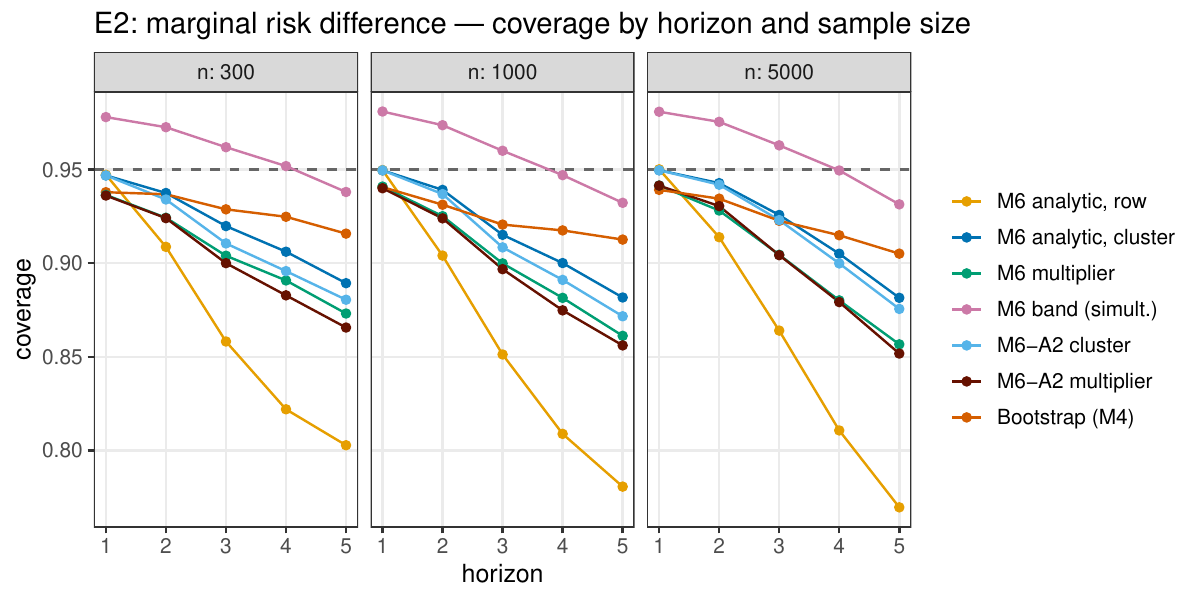}
\caption{E2 (marginal risk difference): coverage by method and horizon, in panels by
sample size. Coverage declines with the horizon for the pointwise methods; the M6
cluster-robust analytic and multiplier intervals track the nonparametric bootstrap,
the row-level analytic interval under-covers most, and the simultaneous band
over-covers pointwise (its simultaneous coverage of the whole curve being $0.88$,
Section~\ref{results}).}
\label{fig:e2}
\end{figure}

\begin{figure}[!tbp]\centering
\includegraphics[width=.72\linewidth,alt={A risk-difference curve over horizons one to five with a shaded pointwise 95 percent multiplier interval and a wider simultaneous sup-t confidence band enclosing it, shown against the true curve which lies within both.}]{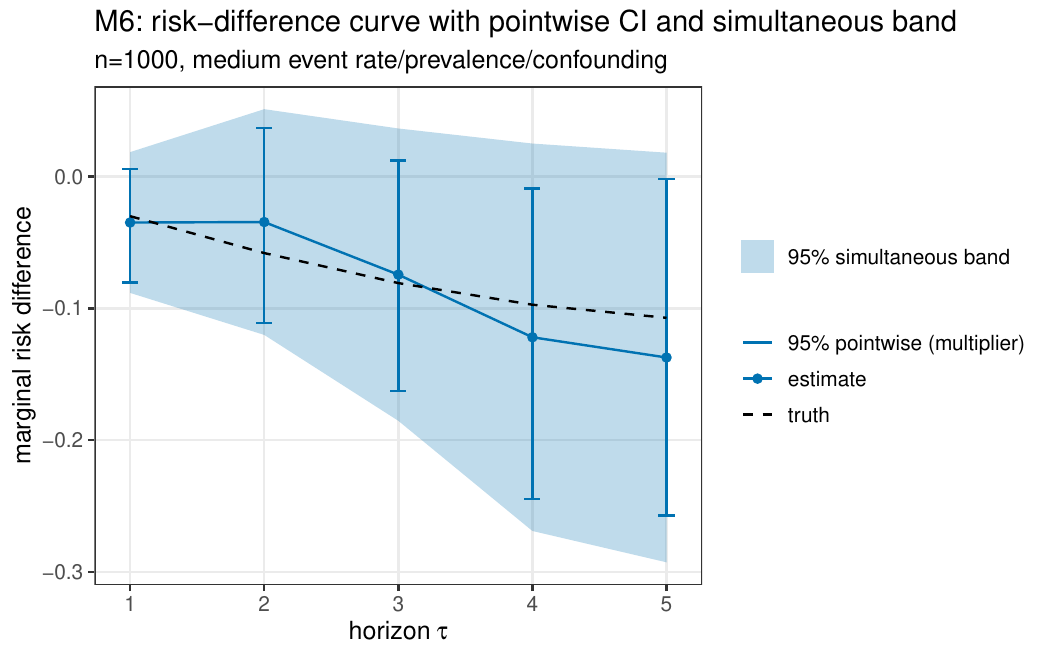}
\caption{Example marginal risk-difference curve ($n=1000$) with the $95\%$
pointwise multiplier interval and the $95\%$ simultaneous (sup-$t$) confidence
band from M6, against the true curve.}
\label{fig:band}
\end{figure}

For the marginal risk-difference curve we compare, at each of the five horizons,
the nonparametric bootstrap (M4) with the four M6 intervals: the analytic
delta-method SE in row-level and cluster-robust form, the multiplier bootstrap,
and the simultaneous band. Table~\ref{tab:e2} and Figure~\ref{fig:e2} give coverage
averaged over horizons and scenarios, and Figure~\ref{fig:band} shows an example
risk-difference curve with its pointwise interval and simultaneous band. The
pattern echoes E1: the row-level analytic interval
under-covers (about $0.86$), whereas the cluster-robust analytic interval
($0.92$) and the multiplier bootstrap ($0.90$--$0.91$) track the nonparametric
bootstrap ($0.92$--$0.93$) closely, their aggregate flat across sample
size---though, as for E1, this aggregate masks a confounding-strength split, the
cluster-robust SE ratio at $\tau=5$ improving with $n$ under weak confounding (to
about $0.96$) but deteriorating under strong confounding (to about $0.68$). This E2 decline, like
its E1 analogue, is a finite-sample effect rather than the proven M1/M2 inconsistency:
consistency of $V_{\mathrm{cl}}$ for E2 (Proposition~1) requires the ratio to return
toward one as $n\to\infty$, the slow approach reflecting the heavy-tailed stabilised
weights, and the omitted weight-estimation term is far too small to account for it
(Approach-2 moves coverage by under half a percentage point). Coverage falls
with the horizon for every pointwise method, and the standard errors are why:
their ratio to the empirical SD is $\approx1$ at $\tau=1$---where a subject's trial
intervals are disjoint---and declines thereafter, to $0.63$ at $\tau=5$ for the
row-level SE, to about $0.80$ for the cluster-robust and multiplier methods, and
least for the bootstrap (Web Table~S8). The simultaneous
band over-covers pointwise on average (about $0.96$), as a band controlling the
whole curve should, although its pointwise coverage falls from $0.980$ at $\tau=1$
to $0.934$ at $\tau=5$ and so does not exceed nominal at every horizon. Its
\emph{simultaneous} coverage---the fraction of replicates in which the band
contains the true curve at all five horizons at once---is $0.881$ ($0.883$,
$0.878$ and $0.882$ at $n=300$, $1000$ and $5000$; $80{,}718$ curves), short of the
nominal $0.95$ and flat in sample size. The shortfall is systematic rather than
driven by a few cells: across the $81$ scenarios simultaneous coverage runs from
$0.754$ to $0.956$ with median $0.887$, and only $3$ scenarios---all at weak
confounding---reach $0.95$. It is nonetheless the only construction considered
here that targets the whole curve, and much the closest to nominal: the pointwise
methods attain simultaneous coverage of only $0.65$--$0.80$. Adding
weight-estimation uncertainty (Approach-2) moves the coverage by at most a few
tenths of a percentage point (Table~\ref{tab:e2}), consistent with the small
change in the standard errors documented below. A small residual
under-coverage persists for the pointwise methods, reflecting variance rather than
the target-definition offset seen at E1. Averaged over scenarios,
coverage also declines with the horizon---for the multiplier interval from
$0.940$ at $\tau=1$ to $0.864$ at $\tau=5$, with the full by-horizon breakdown
for all nine procedures in Web Table~S3---mirroring the pattern
\citet{limozin2025} reported for pooled-logistic STE.

\begin{table}[!htbp]\centering
\caption{Coverage for the marginal risk difference (E2), averaged over the five
horizons and the $27$ scenarios at each sample size; nominal $0.95$. The indented
rows add weight-estimation uncertainty (M6 Approach-2) to the corresponding
Approach-1 method. Computed from the paired per-replicate averages, the
Monte-Carlo standard error of each averaged entry is at most $0.0021$. The band row
reports the \emph{pointwise} coverage of the sup-$t$ band averaged over horizons; its
\emph{simultaneous} coverage of the whole curve is $0.881$ (\S\ref{coverage-e2}), so its
higher pointwise entries do not indicate a better whole-curve interval.}
\label{tab:e2}
\begin{tabular}{lccc}
\toprule
Method & $n=300$ & $n=1000$ & $n=5000$\\
\midrule
M6 analytic, row-level                & 0.868 & 0.859 & 0.862\\
M6 analytic, cluster-robust           & 0.920 & 0.917 & 0.921\\
M6 multiplier bootstrap               & 0.906 & 0.902 & 0.902\\
M6 band, pointwise coverage           & 0.961 & 0.959 & 0.960\\
\;\; Approach-2, analytic cluster-robust & 0.914 & 0.912 & 0.918\\
\;\; Approach-2, multiplier bootstrap    & 0.902 & 0.898 & 0.901\\
M4 nonparametric bootstrap            & 0.929 & 0.925 & 0.923\\
\bottomrule
\end{tabular}
\end{table}

\subsection{Computational cost}\label{computational-cost}
Every method first requires a single prepare-plus-fit of the weighted-Aalen
model---trial construction, weight estimation and the fit---about $23$ s at $n=5000$
on one core; the interval methods then differ only in how many times they refit that
pipeline (Table~\ref{tab:cost}). The analytic standard errors (M1--M3) refit nothing,
being closed-form read-offs of the one fit, and the multiplier/LEF method (M6) also
refits nothing, forming its resampled statistics as linear combinations of
pre-computed influence functions (about $13$ s per replicate). The nonparametric
bootstrap (M4) instead refits the whole pipeline in each of $B=200$ resamples (about
$77$ min per replicate), and the delete-one-subject jackknife (M5) would refit once
per subject---$O(n)$ times---which is why it is evaluated only on the applied example
(Table~\ref{tab:heart}) and not in the main simulation. The multiplier/LEF method thus
attains coverage within one to two percentage points of the bootstrap's
(Table~\ref{tab:e2}), and additionally furnishes simultaneous bands, at a small
fraction of its cost---mirroring, for additive hazards, the efficiency advantage
\citet{limozin2025} reported for the LEF bootstrap under the pooled-logistic model.

\begin{table}[!htbp]\centering
\caption{Per-replicate computational cost of the interval methods at $n=5000$, ordered
by method. Every replicate begins with one \emph{prepare-plus-fit} of the
weighted-Aalen model---trial construction, weight estimation and the fit, about
$23$\,s on one core---which all methods share; the ``refits'' column counts the
\emph{additional} full prepare-plus-fits performed beyond that shared fit: none for
the analytic SEs (M1--M3, closed-form read-offs of the shared fit) or the
multiplier/LEF method (M6, linear combinations of pre-computed influence functions),
$B$ per replicate for the bootstrap (M4), and one per subject for the jackknife (M5).
Wall-clock figures are single-core, order-of-magnitude timings, not a formal
benchmark: M4 and M6 are measured, M1--M3 negligible, and the M5 figure is estimated
as $n\times23$\,s (M5 is run only on the applied example, Table~\ref{tab:heart}). Here
$n$ is the number of subjects.}
\label{tab:cost}
\begin{tabular}{@{}>{\raggedright\arraybackslash}p{0.33\textwidth} >{\centering\arraybackslash}p{0.17\textwidth} >{\raggedright\arraybackslash}p{0.40\textwidth}@{}}
\toprule
Method & Refits per replicate & Wall-clock per replicate ($n=5000$) \\
\midrule
M1--M3 (analytic SEs) & $0$ & negligible beyond the shared $\approx23$\,s fit \\
\addlinespace
M4 (nonparametric bootstrap) & $B=200$ & $\approx77$\,min $\;(=B\times23$\,s$)$ (measured) \\
\addlinespace
M5 (delete-one-subject jackknife) & $n=5{,}000$ & $\approx32$\,h $\;(=n\times23$\,s$)$ (estimated, not run) \\
\addlinespace
M6 (multiplier / LEF) & $0$ & $\approx13$\,s (measured) \\
\bottomrule
\end{tabular}
\end{table}

\section{Application: the Stanford Heart Transplant Data}\label{application}
We illustrate the methods on the Stanford heart transplant study
\citep{crowley1977}, distributed in the R \texttt{survival} package
(\texttt{jasa}) and the canonical public example of a treatment defined at a
time-varying moment. Of $103$ patients accepted for transplantation ($102$ with
positive follow-up), $69$ received a transplant---an absorbing, time-varying
treatment initiated a median of $25$ days after acceptance---and $55$ died within
six months. Unlike the restricted cohorts used by \citet{keogh2023} (UK cystic
fibrosis registry) and \citet{limozin2025} (HERS), these data are fully public,
so the entire analysis is reproducible (Data and code availability).

We discretised follow-up into $30$-day visits over a six-month horizon and
emulated a trial at each visit among patients not yet transplanted, contrasting
``transplant from this visit onward'' with ``never transplant.'' A transplant
occurring within a visit is coded as untreated until the following visit, so the
target is a discrete-visit intervention on the $30$-day grid. Baseline age
(standardised) served as the adjustment covariate; the estimands are the constant
hazard difference of transplant and the marginal risk-difference (in death) curve
at one to six months, with all six confidence-interval methods applied ($B=1000$
resamples for the bootstrap and multiplier methods). Departures from the
trial-baseline treatment were handled by artificial censoring, corrected with
stabilised inverse-probability-of-artificial-censoring weights: a denominator
logistic model for remaining untreated given the standardised age at the next
visit, stabilised by a numerator model in visit number and baseline age. The
emulation produced $246$ person-trials and $569$ analysis rows, on which the
weights were benign---maximum $2.42$, effective sample size $98\%$---and were left
untruncated. The sole adjustment covariate, baseline age, is time-fixed,
so the weights here correct selection into remaining untreated rather than
time-varying confounding affected by prior treatment. The example therefore
exercises the immortal-time-correct comparison and the clustering and
standard-error machinery on real data, rather than the confounding-control role
of the weights.

The point estimates suggest a protective direction---a constant hazard difference
of $-0.037$ per visit and a six-month risk difference of about $-0.09$---but the
uncertainty is wide: every interval comfortably includes zero (the hazard
difference lies about $0.8$ cluster-robust SEs from the null, the six-month risk
difference under one bootstrap SE), and the risk-difference curve is non-monotone
across the six months, as befits $55$ deaths. We therefore read the example as a
standard-error comparison rather than an estimate of a transplant effect. The
methodological pattern seen in simulation nonetheless reappears
(Table~\ref{tab:heart}): the model-based and row-level robust standard errors for
the hazard difference ($0.035$ and $0.034$) are roughly a quarter to a third
smaller than the cluster-robust ($0.046$; $24$--$26\%$), jackknife ($0.048$;
$27$--$29\%$) and bootstrap ($0.051$; $31$--$33\%$) standard errors---a comparison
that does not depend on statistical significance. For the risk-difference curve
(Table~\ref{tab:heart_mrd}), the M6 analytic cluster-robust and
multiplier standard errors track the nonparametric bootstrap at every horizon
(e.g.\ at six months $0.129$ and $0.126$ versus $0.143$), whereas the row-level
analytic SE is materially smaller ($0.092$) and increasingly so as the horizon
grows, consistent on real data with the finding that ignoring the reuse of
subjects across trials understates uncertainty.

To check that the $30$-day discretisation does not drive these conclusions, we
refit the example across visit widths from $30$ down to $7$ days, holding the
six-month horizon fixed ($K=6$ to $26$ visits; Web Table~S9).
The point estimate is grid-dependent: the standardised six-month risk difference
attenuates monotonically from $-0.09$ toward the null (about $+0.02$ at weekly
visits, its interval still covering zero), consistent with coarse intervals
mislabelling a within-interval transplant recipient as untreated and charging an
early death to the untreated arm---a treatment-timing effect, not weight
breakdown, since the stabilised weights stay well behaved throughout (effective
sample size $96$--$98\%$, maximum weight below $3.6$); this reinforces that the
example compares standard errors rather than estimating a transplant effect. The
row-versus-cluster understatement the example exists to show is, by contrast,
invariant to the grid and if anything widens---the row-level robust SE lies
$26\%$ below the subject-clustered SE at $30$-day visits and $45\%$ below at
weekly, corroborated at the finest grid by the delete-one-subject jackknife and
the nonparametric bootstrap, which agree with the analytic clustered SE---as each
subject contributes more correlated trial-rows that the row-level sandwich treats
as independent.

\begin{table}[!htbp]\centering
\caption{Stanford heart transplant: standard error of the constant hazard
difference of transplant ($\hat\gamma=-0.037$ per $30$-day visit) by method.}
\label{tab:heart}
\begin{tabular}{lccccc}
\toprule
 & M1 model & M2 robust & M3 cluster & M5 jackknife & M4 bootstrap\\
\midrule
SE & 0.035 & 0.034 & 0.046 & 0.048 & 0.051\\
\bottomrule
\end{tabular}
\end{table}

\begin{table}[!htbp]\centering
\caption{Stanford heart transplant: marginal risk-difference (of death) estimate and
standard error by horizon and method (E2). The row-level analytic standard error is
increasingly too small, relative to the subject-clustered analytic, multiplier (M6)
and nonparametric bootstrap ($B=1000$) standard errors, as the horizon grows---the
pattern seen in simulation. The fitted cumulative hazard is flat over the final
$30$-day window (no death occurs there), so the five- and six-month rows coincide.}
\label{tab:heart_mrd}
\begin{tabular}{lrcccc}
\toprule
Horizon & MRD & \multicolumn{4}{c}{Standard error}\\
\cmidrule(l){3-6}
(months) & estimate & row-level & cluster-robust & multiplier & bootstrap\\
\midrule
1 & $-0.051$ & 0.053 & 0.053 & 0.054 & 0.053\\
2 & $\phantom{-}0.009$ & 0.071 & 0.080 & 0.080 & 0.084\\
3 & $-0.071$ & 0.077 & 0.095 & 0.094 & 0.102\\
4 & $-0.119$ & 0.081 & 0.109 & 0.108 & 0.119\\
5 & $-0.090$ & 0.092 & 0.129 & 0.126 & 0.143\\
6 & $-0.090$ & 0.092 & 0.129 & 0.126 & 0.143\\
\bottomrule
\end{tabular}
\end{table}

\section{Discussion}\label{discussion}

\subsection{Summary of findings}\label{summary-of-findings}
\textbf{The failure.} Under a correctly specified additive-hazards marginal structural model, the
sequential-trials estimator's default standard errors---the model-based SE and
the row-level robust (sandwich) SE reported by standard software
(\texttt{timereg::aalen})---are markedly anticonservative, and the shortfall
does not diminish as the sample grows. In the simulation the coverage of nominal
$95\%$ intervals for the constant hazard difference falls from about $0.82$ at
$n=300$ to $0.71$--$0.72$ at $n=5000$, with the ratio of mean estimated SE to
empirical standard deviation well below $1$ throughout---about $0.70$ at $n=300$
falling to about $0.60$ at $n=5000$ (Table~\ref{tab:e1}).
A proportional SE deficit that persists as $n\to\infty$ is the signature of an
inconsistent, rather than a merely finite-sample, variance estimator---an
inconsistency Proposition~2 establishes for the row-level robust SE, the
model-based SE being anticonservative as an empirical finding
(Table~\ref{tab:roadmap}). The
mechanism is that each subject contributes to as many as five overlapping
emulated trials, so the analysis records are strongly correlated, yet the
row-level estimators treat them as independent. Two opposing forces act on the naive
variance---ignoring the estimation of the weights inflates it \citep{austin2016},
while ignoring the reuse of subjects deflates it---and, empirically, the
subject-reuse effect dominates by a wide margin.

\textbf{The remedy for the hazard difference.} Clustering
the robust sandwich variance on the subject (M3) is cheap and effective: it addresses exactly the offending
dependence and recovers most of the deficit, lifting coverage to about $0.91$,
$0.90$ and $0.86$ across the three sample sizes with interval widths essentially
identical to the nonparametric bootstrap; the shortfall that remains at $n=5000$
is predominantly the target-definition offset of
\S\ref{bias-of-the-point-estimator} rather than a variance failure. Because the additive hazard difference
is collapsible over baseline covariates, this scalar target is a tractable core:
no non-collapsibility bias arises in targeting it, and a single clustered SE
from off-the-shelf software suffices (the sustained-treatment marginal effect
still differs from the per-visit conditional coefficient $\alpha_A$ because $L$ is
treatment-affected, as noted in Section~\ref{results}).

\textbf{The remedy for the risk-difference curve.} This estimand is of primary
substantive interest \citep{keogh2023}; we adapted the linearised-estimating-function
(LEF) bootstrap of \citet{limozin2025} to the closed-form weighted-Aalen
estimator \citep{aalen1989}. Because that estimator is weighted least squares, the influence function is
available in closed form with no Newton step (the estimator is asymptotically,
not exactly, linear---see the Remark in Section~\ref{sec:theory}), and the LEF
bootstrap reduces to a fast multiplier (wild) bootstrap of pre-computed influence
functions, with no refitting. The resulting
cluster-robust analytic and multiplier intervals track the nonparametric
bootstrap closely and stably across sample size (coverage near $0.90$--$0.92$
versus $0.92$--$0.93$; Table~\ref{tab:e2}), whereas the row-level analytic
interval again under-covers---confirming that the correct clustering unit is the
subject, not the individual analysis record. The method additionally supplies simultaneous
confidence bands for the whole MRD curve, which over-cover pointwise (about
$0.96$), as a band controlling the whole curve should; their \emph{simultaneous}
coverage is nonetheless $0.881$ rather than the nominal $0.95$
(\S\ref{coverage-e2}). The multiplier approach attains this at a fraction of the bootstrap's
cost: it computes the influence functions once (about $13$~s at $n=5000$) and
forms each resampled statistic as a linear combination, against the bootstrap's
$B$ full refits of trial construction, weight estimation and the weighted Aalen
fit (about $23$~s each). This mirrors, for additive hazards, the efficiency
advantage \citet{limozin2025} reported for the pooled-logistic MSM.

\subsection{Weight-estimation and standardisation uncertainty are negligible}\label{weight-negligible}
The multiplier bootstrap makes two approximations relative to the nonparametric
bootstrap---it conditions on the estimated weights and on the empirical
standardisation population---and we ruled out both as sources of the residual SE
shortfall. First, the \emph{weights}. The Approach-2 variant propagates
weight-estimation uncertainty analytically---differentiating the cumulative
stabilised weights through the two logistic weight models and adding the resulting
influence-function correction (Web Appendix~B)---and it changes the
marginal-risk-difference standard errors by a signed median of $-0.35\%$, exceeding
$1\%$ in magnitude
in $36\%$ of design cells and reaching $4.7\%$ at the longest horizons: with
\emph{stabilised} weights the correction is close to orthogonal to the outcome
influence function \citep{shu2021,enders2018}. It is far too small, and of the
wrong sign in most cells, to account for the shortfall. The constant hazard
difference (E1) is fitted with the \emph{same} stabilised weights---about $71\%$ of
analysis rows carry weight exactly one for structural reasons---so the same
conclusion applies to it, and we do not compute a separate E1 correction. Second, the
\emph{standardisation population}. Propagating its sampling variability adds a
g-computation empirical-process term $\{S_0^{(i)}-S_1^{(i)}-\widehat{\mathrm{MRD}}\}/n$
to each subject's influence function (Web Appendix~A; the random-standardisation
option of \texttt{steCI}). This term is $O(1/n)$---the same order
as every other component, but $\widehat{\mathrm{MRD}}$ is a sample mean, so its
contribution is the small $\mathrm{Var}\{S_0^{(i)}-S_1^{(i)}\}/n$---and moving the
standardisation from fixed to random changes the standard errors and coverage
negligibly (Web Appendix~D), matching the nonparametric bootstrap that resamples
that population. Both secondary uncertainties are therefore negligible with
stabilised weights, so the fixed-weight, fixed-standardisation Approach-1 suffices;
the Approach-2 and random-standardisation machinery are worth carrying mainly as
the checks that establish this. What this eliminates, however, is two specific
\emph{variance corrections}---not the variance explanation itself---and the residual
is attributable only for one of the two estimands. For the constant hazard
difference, recentring the intervals on the mean estimate recovers about half the
shortfall at $n=300$ and $1000$ and most of it ($79\%$) at $n=5000$---by $0.019$ to
$0.075$ across sample sizes, lifting cluster-robust
coverage to $0.923$--$0.930$; there the residual is largely the
\emph{target-definition offset} of \S\ref{bias-of-the-point-estimator}---the estimator's
front-loaded weighting of time since initiation, applied to a declining hazard
difference---which erodes coverage once the standard error is small. It is not a
defect of the estimator, whose intervals are correctly centred on their own
estimand; it is the price of scoring them against a differently weighted target.
For the risk-difference curve it is not: on the same per-sample-size basis,
recentring buys at most $0.009$---against $0.019$ to $0.075$ for E1---and leaves
cluster-robust coverage at $0.924$ against a nominal $0.95$. That remaining gap is predominantly a \emph{variance} effect which neither
omitted correction explains, and which skewness- and bias-corrected bootstrap
intervals do not close either; we quantify it as a limitation rather than attribute
it further (\S\ref{limitations}).

\subsection{Practical guidance for analysts}\label{practical-guidance}
For applied users of the additive-hazards sequential-trials estimator we suggest
the following defaults.
\begin{enumerate}
\item[(i)] The model-based and the row-level robust standard errors from the
outcome model should not be reported: both ignore the reuse of subjects
across stacked trials and are severely anticonservative, the more so at larger
$n$.
\item[(ii)] For the constant hazard difference, report the cluster-robust sandwich
SE with clustering on the subject identifier; it is available directly from
standard software, costs nothing extra, and recovers most of the coverage
deficit. Bear in mind that the constant hazard difference so estimated is a
design-weighted summary of a possibly time-varying effect
(\S\ref{bias-of-the-point-estimator}), dependent on the emulated-trial structure,
rather than a transportable single-cohort causal parameter.
\item[(iii)] For the marginal risk-difference curve, report the multiplier/LEF
intervals (cluster-robust analytic or multiplier-bootstrap), and accompany the
pointwise intervals with the simultaneous confidence band whenever the curve is
interpreted as a whole rather than at a single pre-specified horizon. This band is
simultaneous over the pre-specified horizon grid; its measured simultaneous coverage
in our study is about $0.88$ rather than the nominal $0.95$ (\S\ref{coverage-e2}), so
it is best read as the closest-to-nominal of the whole-curve constructions we compare
rather than as an exact $95\%$ band.
\item[(iv)] Treat the nonparametric bootstrap of \citet{keogh2023} as the full-refit
comparator: it relaxes both the fixed-weight and the fixed-standardisation-population
assumptions, and the analytic and multiplier methods should be read as fast,
band-supplying approximations to it rather than as replacements when computation
is not a constraint.
\item[(v)] The additive model does not constrain the fitted hazard to
be non-negative, and neither our implementation nor \texttt{timereg} floors it, so
report the fraction of fitted increments of the cumulative baseline that are
negative---directly computable from the fitted object---which lets readers judge how
close the analysis sits to the degenerate low-event-rate boundary at which the
additive specification stops being tenable.
\item[(vi)] The
fixed-weight (Approach-1) variance is adequate for practical purposes here:
propagating weight-estimation uncertainty changed our standard errors by a signed median of
$-0.35\%$ (up to $4.7\%$ in magnitude at the longest horizons). We would not generalise this beyond
sustained or absorbing strategies, however: about $71\%$ of our analysis rows carry
weight exactly one for \emph{structural} reasons---the treated arm by construction
once treatment is absorbing, and every trial's first visit, whose cumulative product
is empty---leaving under a third of rows on which the weight model can act at all.
A design with non-absorbing treatment and longer control follow-up would expose more
of the analysis to weight estimation, and the correction should be checked rather
than assumed negligible.
\end{enumerate}

\subsection{Limitations}\label{limitations}
Several assumptions frame these conclusions. First, identification of the causal
estimands rests on no unmeasured confounding, correct specification of the
\emph{denominator} weight model, and correct specification of the additive marginal
structural model; our data-generating mechanism satisfies the first two by
construction \citep{keogh2021sim}, and the third up to a small residual for the
cumulative-coefficient (E2) target. The mechanism makes an additive marginal
structural model correctly \emph{specifiable}, but our fit adjusts on
trial-baseline covariates only and pools $B(\cdot)$ across origins, and
Web Appendix~E shows this leaves an asymptotic bias of order $0.004$ at $\tau=5$
in the worst cell---of which the pooling component is at most about $0.001$, as
detailed below. The
stabilising numerator, pooled across trial origins with no origin term, is a working
model and is not correctly specified---the same origin pooling that makes $B(\cdot)$
a working convention leaves the adherence probability given trial-baseline covariates
mildly origin-dependent---but it is immaterial to the estimating equation: the
numerator is a predictable function of covariates already in the outcome model, so it
rescales the estimating function without disturbing its mean-zero property. The
constant-effect E1 specialisation is a working summary of a marginal hazard
difference that varies by a factor of $2.1$ over follow-up, which is the source of
the E1 target-definition offset (Section~\ref{bias-of-the-point-estimator}). This is what largely isolates
variance-estimation performance from model misspecification; but it also means
our coverage results speak predominantly to the variance problem---save for the
small, localised misspecification bias documented below---and not to robustness
against a grossly wrong outcome or weight model.

Second, the additive-hazard
non-negativity requirement is intrinsic, not incidental: because the treated
arm's confounder drifts downward, the fitted hazard turns negative at low
baseline hazards, so the achievable low-event-rate range is bounded from below
(at $\alpha_0=0.05$ roughly $44\%$ of hazard evaluations are truncated, against a
negligible fraction at $\alpha_0\ge0.15$). Under this data-generating mechanism,
additive-hazards STE therefore cannot probe the very-low-event regime that a
pooled-logistic MSM can, and comparisons
with \citet{limozin2025} must be read with that boundary in mind.

Third, a
residual under-coverage remains for all pointwise methods, including the
nonparametric bootstrap, and it \emph{grows} with the horizon rather than
subsiding: coverage of the multiplier interval falls from $0.940$ at $\tau=1$ to
$0.864$ at $\tau=5$ and of the cluster-robust analytic interval from $0.949$ to
$0.884$, while the nonparametric bootstrap declines more gently ($0.939$ to
$0.911$; Web Table~S3). For the risk-difference curve this residual is a \emph{variance} effect
rather than point-estimator bias: averaged over scenarios, recentring the intervals
on the mean estimate raises coverage by at most $0.009$ at any sample size, against
$0.019$--$0.075$ for the constant
hazard difference, where the target-definition offset
(\S\ref{bias-of-the-point-estimator}) accounts for it. That average conceals one
corner, and we flag it rather than let it be averaged away: at $n=5000$ under strong
confounding and the low event rate, recentring at $\tau=5$ lifts coverage by
$0.05$--$0.07$ depending on method. There, and only there, the E2 residual is also
substantially a bias effect, and it does not fully subside as $n$ grows. A
population-limit calculation for this cell (Web Appendix~E) attributes the small asymptotic
floor---of order $0.004$ at $\tau=5$, against a curve near $-0.18$---to
outcome-model misspecification: the additive fit adjusts on trial-baseline
covariates, while the hazard depends on the time-varying confounder $L_k$ and the
baseline frailty, and an ablation that removes the outcome's dependence on those
two collapses the bias to zero. The origin pooling of $B(\cdot)$, the natural
suspect, contributes at most about $0.001$ at every horizon (Web Appendix~E); no
new pooling bias accrues at the longest horizons, where the weight mass comes from
a single trial origin. The finite-sample bias in this corner is larger (about $0.025$ at
$n=5000$) and declines with $n$ toward this asymptotic floor, so the corner
reflects a genuine if minor misspecification, not a finite-sample artefact alone. The two variance corrections the multiplier bootstrap omits
are both far too small to account for it (\S\ref{weight-negligible}), and skewness- and
bias-corrected (BCa) bootstrap intervals gave no indication of closing it either:
in a small two-cell probe at $n=300$ in which the BCa and percentile limits were
read off the \emph{same} bootstrap distribution---so the comparison is paired by
construction and differs only in the quantile levels taken---BCa did not improve on
the percentile interval. The probe is small and we do not treat interval shape as
formally excluded. We
therefore report the shortfall as a quantified limitation rather than as a
diagnosed defect.

Fourth, the
primary data-generating mechanism carries a single time-varying confounder. The
anticonservatism of the row-level standard error is, however, a \emph{structural}
consequence of subject reuse across trials (Proposition~2) rather than a feature
of any particular mechanism, and three further data-generating mechanisms
(Web Appendix~D)---a second time-varying confounder, a \emph{misspecified}
(Cox-type) outcome model, and a time-varying treatment effect---confirm that the
row-under-covers, cluster-corrects pattern and the positive sign of the
cross-covariance share (C5) are unchanged. We do not vary the monotone (once-on,
always-on) treatment: sequential trial emulation targets \emph{sustained}
treatment strategies, for which an absorbing treatment is the defining setting
\citep{keogh2023,gran2010}; non-monotone or dynamic treatment regimes are a
different estimand requiring a different estimator, and lie outside our scope. All
results are from the full design of $81$ scenarios, targeting $1000$ replicates
per cell, of which $80{,}988$ of $81{,}000$ yield a point estimate---one cell
falls short, at $988$---while interval construction fails additionally and
unevenly across methods: the worst-affected cell, taken across all nine interval
methods, retains $793$, while the multiplier's own worst cell ($812$) is the one
quantified at the head of Section~\ref{results}. Because each method is then scored on a slightly
different subset, we checked that the exclusion is not driving the comparison:
$99.6\%$ of replicate--horizon combinations yield all nine risk-difference intervals,
and re-scoring every method on that common subset changes coverage by at most
$0.0018$ (the basic nonparametric bootstrap at $n=300$; at most $0.0007$ for the
seven methods reported in Table~\ref{tab:e2}) and by less elsewhere. The exclusion is
informative---dropped replicates are the extreme ones---but bounded well below the
differences we interpret. The Monte-Carlo error of the
sample-size-averaged coverages in
Tables~\ref{tab:e1}--\ref{tab:e2} (each an average over $27$ scenarios) is small
(computed from the paired per-replicate averages, at most $0.0043$ for
Table~\ref{tab:e1} and $0.0021$ for Table~\ref{tab:e2}; replicate $i$ uses a common
seed stream across scenarios, so the per-scenario coverage indicators are positively
correlated and the independence bound of $0.0028$ understates the former by about a
third), while individual-cell coverages carry larger
errors (at most $0.016$, and about $0.0095$ where coverage is near $0.9$) and
the $300$-replicate two-confounder robustness cells of
Web Appendix~D about $0.02$--$0.03$ \citep{morris2019}, so the headline
conclusions are not attributable to simulation noise. The common seed stream cuts the
other way for the comparisons we actually draw: because every method is evaluated on
the same replicates, the \emph{differences} between methods are paired, and their
Monte-Carlo errors are correspondingly smaller than these marginal bounds suggest---at
most about $0.0026$ for Table~\ref{tab:e1}, so differences of roughly $0.005$ or more
are outside simulation noise.

\subsection{Extensions}\label{extensions}
Four extensions follow naturally. First, our design uses administrative censoring
only; incorporating inverse-probability-of-censoring weighting for dependent
(informative) censoring, and propagating its estimation into the influence
functions alongside the artificial-censoring weights, is a direct generalisation
of the Approach-2 machinery. Second, many applications feature competing risks;
an additive-hazards STE for the subdistribution (or cause-specific cumulative
incidence), with matching multiplier-bootstrap bands, would extend the
risk-difference target to that setting. Third, the point-estimator bias of the
standardised risk difference in the strong-confounding low-event-rate corner
identified in \S\ref{limitations} invites a targeted bias correction---a
one-step or targeted-minimum-loss update of the cumulative coefficients, for which
the influence-function framework already supplies the ingredients---though no such
update touches the E1 target-definition offset of
\S\ref{bias-of-the-point-estimator}, whose constructive response (declaring the
weighting, or reporting the time-varying hazard difference directly rather than a
constant summary) is set out there. Fourth, the
weight models here are low-dimensional, with a correctly specified denominator and a
numerator that need only be a working model; data-adaptive (machine-learning) estimation of the weights,
with the attendant need for cross-fitting to preserve valid inference, is a
practically important direction, and the negligible weight-estimation correction
we document with stabilised weights is a useful baseline against which to judge
whether such flexibility changes the variance in a way that must be accounted
for.

\subsection{Conclusion}\label{conclusion}
For the weighted-Aalen sequential-trials estimator, the default model-based and
row-level robust standard errors are anticonservative, not merely optimistic in
finite samples, with under-coverage that does not diminish as the sample grows
because both ignore the reuse of each subject across overlapping emulated
trials---an inconsistency Proposition~2 establishes for the row-level robust SE
and that we document empirically for the model-based SE. Clustering the
sandwich variance on the subject removes most of the deficit for the constant hazard
difference at no additional cost, and a fast multiplier/LEF bootstrap with simultaneous
confidence bands does the same for the marginal risk-difference curve, approaching the
nonparametric bootstrap's coverage at a fraction of its cost. A small residual remains,
traceable for the hazard difference to the design's target-definition offset rather
than to a variance failure. We therefore recommend clustering on the subject and
reporting the multiplier/LEF intervals, and against reporting the row-level or
model-based standard errors in this design.

\section*{Supplementary Material}
Web Appendices A--E, referenced throughout, are provided in the accompanying
Supplementary Material: (A) influence functions for the weighted-Aalen STE
estimator and their mapping to the \texttt{steCI} software; (B) the Approach-2
weight-estimation correction; (C) proofs of Propositions~1--3; (D) the
data-generating process, the scenario grid, an additional coverage figure,
and robustness studies; and (E) the asymptotic (population-limit) analysis of
the residual E2 bias and its attribution.

\section*{Data and code availability}
All methods are implemented in the open-source \texttt{steCI} R package. The
package, the full simulation study (scenario grid, cluster job scripts, and
aggregated results), the applied analysis, and the validation and robustness
scripts referenced throughout are available at
\url{https://github.com/ehsanx/steCI}; the Stanford heart transplant data are
public in the R \texttt{survival} package (dataset \texttt{jasa}). All reported
simulation results were produced under R~4.4.0 with \texttt{timereg}~2.0.7.

\section*{Ethics and consent}
This study analyses only the fully de-identified, publicly available Stanford heart
transplant dataset (\texttt{jasa}), distributed in the R \texttt{survival} package; no
new ethical approval or participant consent was required.

\section*{Declaration of conflicting interests}
The author declares that there is no conflict of interest.

\section*{Funding}
The author received no financial support for the research, authorship, or publication
of this article.

\section*{Acknowledgements}
This research was enabled in part by computational resources provided by Advanced
Research Computing at the University of British Columbia. During the preparation of this
work the author used AI-based tools to assist with simulation and analysis code,
manuscript editing, and the checking of derivations; the author reviewed and verified
all outputs and takes full responsibility for the content of this article.

\bibliography{references}
\end{document}